\documentclass[longauth]{aa}  

\usepackage{hyperref}
\hypersetup{
  colorlinks=true,   
  citecolor=blue,    
  urlcolor=blue,     
  linkcolor=blue,     
}
\usepackage{natbib}
\bibpunct{(}{)}{;}{a}{}{,} 
\usepackage{graphicx}
\usepackage{multirow}
\usepackage{txfonts}

\makeatletter
\renewcommand*\aa@pageof{, page \thepage{} of \pageref*{LastPage}}
\makeatother

\begin{document}

   \title{SUPER VIII}

   \subtitle{Fast and Furious at $z\sim2$: obscured type-2 active nuclei host faster\\ ionised winds than type-1 systems}

    \authorrunning{G. Tozzi et al.}
    \titlerunning{SUPER VIII. Ionised outflows faster in type-2 AGN}

    \author{G. Tozzi\inst{1,2,3}\thanks{\email{gtozzi@mpe.mpg.de}},
    G. Cresci\inst{2},
    M. Perna\inst{4},
    V. Mainieri\inst{5},
    F. Mannucci\inst{2},
    A. Marconi\inst{1,2},
    D. Kakkad\inst{6},
    A. Marasco\inst{7},\\
    M. Brusa\inst{8,9},
    E. Bertola\inst{2},
    M. Bischetti\inst{10,11},
    S. Carniani\inst{12},
    C. Cicone\inst{13},
    C. Circosta\inst{14,15},
    F. Fiore\inst{11,16},
    C. Feruglio\inst{11,16},
    C. M. Harrison\inst{17},
    I. Lamperti\inst{4},
    H. Netzer\inst{18},
    E. Piconcelli\inst{19},
    A. Puglisi\inst{20},
    J. Scholtz\inst{21,22},
    G. Vietri\inst{23},\\
    C. Vignali\inst{8,9},
    and G. Zamorani\inst{9}
    }

    \institute{$^{1}$ Dipartimento di Fisica e Astronomia, Università di Firenze, Via G.\ Sansone 1, 50019 Sesto Fiorentino, Firenze, Italy\\
    $^{2}$ INAF - Osservatorio Astrofisico di Arcetri, Largo E.\ Fermi 5, 50125 Firenze, Italy\\
    $^{3}$ Max-Planck-Institut für Extraterrestrische Physik (MPE), Giessenbachstraße 1, D-85748 Garching, Germany\\
    $^{4}$ Centro de Astrobiología (CAB), CSIC–INTA, Cra. de Ajalvir Km. 4, 28850 – Torrejón de Ardoz,
    Madrid, Spain\\
    $^{5}$ European Southern Observatory, Karl-Schwarzschild-Strasse 2, Garching bei München, Germany\\
    $^{6}$ Space Telescope Science Institute, 3700 San Martin Drive, Baltimore, MD 21218, USA\\
    $^{7}$ INAF - Osservatorio Astronomico di Padova, Vicolo dell’Osservatorio 5, I-35122 Padova, Italy\\
    $^{8}$ Dipartimento di Fisica e Astronomia “Augusto Righi”, Università di Bologna, Via P. Gobetti 93/2, 40129, Bologna, Italy\\
    $^{9}$ INAF - Osservatorio di Astrofisica e Scienza dello Spazio, Via Gobetti 93/3, 40129 Bologna, Italy\\
    $^{10}$ Dipartimento di Fisica, Università di Trieste, Sezione di Astronomia, Via G.B. Tiepolo 11, I-34131 Trieste, Italy\\
    $^{11}$ INAF - Osservatorio Astronomico di Trieste, Via G. Tiepolo 11, I-34143 Trieste, Italy\\
    $^{12}$ Scuola Normale Superiore, Piazza dei Cavalieri 7, I-56126 Pisa, Italy\\
    $^{13}$ Institute of Theoretical Astrophysics, University of Oslo, PO Box 1029, Blindern 0315, Oslo, Norway\\
    $^{14}$ Department of Physics \& Astronomy, University College London, Gower Street, London, WC1E 6BT, UK\\
    $^{15}$ ESA, European Space Astronomy Centre (ESAC), Camino Bajo del Castillo s/n, 28692 Villanueva de la Cañada, Madrid, Spain\\
    $^{16}$ IFPU - Institute for fundamental physics of the Universe, Via Beirut 2, 34014 Trieste, Italy\\
    $^{17}$ School of Mathematics, Statistics and Physics, Newcastle University, Newcastle upon Tyne, NE1 7RU, UK\\
    $^{18}$ School of Physics and Astronomy, Tel-Aviv University, Tel Aviv 69978, Israel\\
    $^{19}$ INAF - Osservatorio Astronomico di Roma, Via di Frascati 33, 00040, Monteporzio Catone, Rome, Italy\\
    $^{20}$ School of Physics and Astronomy, University of Southampton, Highfield SO17 1BJ, UK\\
    $^{21}$ Kavli Institute for Cosmology, University of Cambridge, Madingley Road, Cambridge, CB3 0HA, UK\\
    $^{22}$ Cavendish Laboratory - Astrophysics Group, University of Cambridge, 19 JJ Thompson Avenue, Cambridge, CB3 0HE, UK\\
    $^{23}$ INAF – Istituto di Astrofisica Spaziale e Fisica Cosmica Milano, Via A. Corti 12, 20133 Milano, Italy\\
    }

   \date{Received XX; accepted XX}

 
  \abstract
   {We present spatially resolved VLT/SINFONI spectroscopy with adaptive optics of type-2 active galactic nuclei (AGN) from the SINFONI Survey for Unveiling the Physics and Effect of Radiative feedback (SUPER), which targeted X-ray bright ($L_{\rm 2-10~keV}\gtrsim10^{42}$ erg s$^{-1}$) AGN at the Cosmic Noon ($z\sim2$). Our analysis of the rest-frame optical spectra unveils ionised outflows in all seven examined targets, as traced via {\normalfont [O\,{\sc iii}]}$\lambda$5007 line emission, moving at $v\gtrsim600$ km s$^{-1}$. In six objects these outflows are clearly spatially resolved and extend on 2--4 kpc scales, whereas marginally resolved in the remaining one. Interestingly, these SUPER type-2 AGN are all heavily obscured sources ($N_{\rm H}\gtrsim10^{23}$ cm$^{-2}$) and host faster ionised outflows than their type-1 counterparts within the same range of bolometric luminosity ($L_{\rm bol} \sim 10^{44.8-46.5}$ erg s$^{-1}$). SUPER has hence provided observational evidence that the type-1/type-2 dichotomy at $z\sim2$ might not be driven simply by projection effects, but might reflect two distinct obscuring life stages of active galaxies, as predicted by evolutionary models. Within this picture, SUPER type-2 AGN might be undergoing the ‘blow-out' phase, where the large amount of obscuring material efficiently accelerates large-scale outflows via radiation pressure on dust, eventually unveiling the central active nucleus and signal the start of the bright, unobscured type-1 AGN phase. Moreover, the overall population of ionised outflows detected in SUPER has velocities comparable with the escape speed of their dark matter halos, and in general high enough to reach 30--50 kpc distances from the centre. These outflows are hence likely to sweep away the gas (at least) out of the baryonic disk and/or to heat the host gas reservoir, thus reducing and possibly quenching star formation.}
   

   \keywords{galaxies: active - galxies: evolution - galaxies: high-redshift - quasars: emission lines - techniques: imaging spectroscopy}

   \maketitle
%
\section{Introduction}

A key question in galaxy evolution is how active galactic nuclei (AGN) interact with their host galaxy and shape its physical properties, from gas content to star formation (SF) and chemical enrichment. This so-called AGN feedback typically acts via massive outflows (e.g. \citealt{King2005,Fabian2012,Costa2014}), powered on sub-pc scales by the radiative output of accreting supermassive black holes (BH), and then accelerated up to galaxy scales of $\sim$1--10 kpc (e.g. \citealt{King2015}). These winds are considered the main responsible for quenching SF within their host, by heating or expelling the gas reservoir from which new stars form (i.e. ‘preventive' versus ‘ejective' feedback), and settling the observed scaling relations between BH mass and host galaxy properties (e.g. \citealt{Ferrarese2000,Marconi2003,Kormendy2013,Marasco2021}).

To date we have plenty of observational evidence of AGN-driven outflows from low (e.g. \citealt{Feruglio2010,Harrison2014,Woo2016}) to high redshift (z$>$1; e.g. \citealt{Maiolino2012,Foerster2014,Carniani2023}), in different gas phases (e.g. \citealt{Cicone2018}). While mm/sub-mm interferometric observations are optimally suited to investigate the cold molecular ($T$$<$100 K) component of AGN-driven outflows (typically via CO line emission; \citealt{Cicone2014,Morganti2015,Chartas2020}), optical/near-IR Integral Field Spectroscopy (IFS) enables detailed spatially resolved studies of the warmer ionised component of outflows. To trace ionised outflows, the most commonly adopted tracer is the optical [O\,{\sc iii}]$\lambda$5007 line emission (e.g. \citealt{CanoDiaz2012,Venturi2018,Marshall2023}), a forbidden line transition which can originate only from low-density regions ($n_{\rm e}<10^6$ cm$^{-3}$), thus excluding any potential contamination from high-density AGN Broad Line Region (BLR). Therefore, [O\,{\sc iii}]$\lambda$5007 is an excellent tracer of ionised gas extended on 1--10 kpc scales. 

Thanks to high-quality IFS observations, obtained with ground-based adaptive optics (AO) assisted facilities or from space with the \textit{James Webb Space Telescope}, AGN-driven ionised outflows and feedback mechanisms have been deeply investigated at z$>$1 (e.g. \citealt{Foerster2014,Perna2015b,Perna2023,Wylezalek2022,Cresci2023,Marshall2023,Vayner2023}), with the primary focus on the so-called Cosmic Noon (z$\sim$2), where both SF and BH accretion histories reach the peak of their activity \citep{Madau2014}. This makes the redshift range z$\sim$1--3 the golden epoch of AGN feedback, where its effects are expected to be maximised.

Nonetheless, the high-redshift search has primarily focused so far on brighter distant sources ($L_{\rm bol}>10^{46}$ erg s$^{-1}$; e.g. \citealt{Bischetti2017,Perrotta2019}), easier to detect and to examine, which has led to a biased census of high-redshift AGN-driven outflows. In addition to this, the lack of direct observational evidence confirming (or disproving) key theoretical predictions prevents us from a clear and complete picture of AGN feedback and outflows, with several questions still open. For instance, how do AGN-driven outflows really affect their host galaxy? Do they quench SF activity efficiently? Are they representative of a particular life stage of galaxies' evolution?

One of these theoretical scenarios still waiting for observational proof predicts that AGN galaxies experience a first dust-enshrouded life stage (e.g. \citealt{Hopkins2006,Menci2008}), where the central nucleus is obscured by dust and gas, which fuel both supermassive BH growth and SF efficiently. This sets the stage to the short-lived (a few tens of Myr) ‘blow-out’ phase (e.g. \citealt{Hopkins2008}), during which powerful AGN-driven winds sweep away obscuring material and make the central AGN visible and unobscured. Such an evolutionary scenario was proposed to explain the observed dichotomy between red and blue quasars (e.g. \citealt{Brusa2010,Klindt2019,Perrotta2019,Fawcett2023}), with the former representing the brief obscured, transitional phase featured by powerful winds accelerated via radiation pressure on dust (e.g. \citealt{King2015,Costa2018}). For this reason, X-ray bright (easier to detect) but optically obscured AGN at $z=1-3$ have been deeply investigated for years, as optimal candidates undergoing the blow-out phase (e.g. \citealt{Brusa2015a,Brusa2016,Cresci2015,Cresci2023,Perna2015a,Perna2015b,Veilleux2023}), hence as the most promising targets to catch AGN feedback and outflows in action.

To draw a coherent picture of AGN feedback at the Cosmic Noon, it is fundamental to conduct systematic and unbiased searches for outflows in statistically large samples. The KMOS AGN Survey at High redshift (KASHz; \citealt{Harrison2016}) has provided spatially-resolved information for hundreds of X-ray selected AGN, revealing ionised gas velocities likely indicative of outflows in about 50\% of the examined sample \citep{Harrison2016}. However, KASHz employed seeing-limited observations with a typical spatial resolution of 4--9 kpc at $z>1$ (i.e. $>$0.5$''$), which is not sufficient to trace outflows and their properties in detail. To reach higher spatial resolution, we can exploit more time-consuming AO-assisted observations of smaller samples, yet still representative of the AGN population at $z\sim2$.

Our completed Large Programme SUPER (Survey for Unveiling the Physics and Effect of Radiative feedback; PI: V. Mainieri, ID: 196.A-0377), carried out with the near-IR spectrograph SINFONI \citep{Eisenhauer2003} at the Very Large Telescope (VLT) of the European
Southern Observatory (ESO), has collected AO-assisted IFS data of a representative sample of X-ray AGN ($L_{\rm 2-10~keV}\gtrsim10^{42}$ erg s$^{-1}$) at z$\sim$2, spanning a wide range of bolometric luminosity ($L_{\rm bol}
=10^{44-48}$ erg s$^{-1}$). Presented by \citet{Circosta2018}, the SUPER survey offers us the extraordinary possibility of investigating AGN-driven ionised outflows in a few tens of AGN at the Cosmic Noon, selected in an unbiased way with respect to the chance of hosting outflows. Thanks to SINFONI AO system, we reached exquisite angular resolutions of 0.2$''$--0.5$''$, capable of probing spatial scales down to 2--4 kpc at z$\sim$2. 

In such a perspective, this paper continues the series of publications dedicated to SUPER\footnote{Link to the SUPER webpage \href{http://www.super-survey.org/}{\textcolor{blue}{http://www.super-survey.org/}}.}, by presenting SINFONI observations of the type-2 AGN subsample. In particular, this work complements the analysis carried out by \citet{Kakkad2020} of the ionised outflows in the SUPER type-1 AGN (see also \citealt{Vietri2020}), by extending the study of ionised outflows to the SUPER type-2 systems (see also \citealt{Lamperti2021}). We still employ [O\,{\sc iii}]$\lambda$5007 line emission to trace large-scale ionised outflows, and then compare the inferred properties of the type-2 ionised outflows with those of the type-1 counterparts, aiming at uncovering possible differences between these two AGN populations.

This paper is organised as follows. In Sect. \ref{sec2}, we introduce the SUPER type-2 AGN subsample examined in this work, and present our SINFONI observations as well as data reduction strategy. In Sect. \ref{sec3}, we describe the fitting procedure adopted to derive ionised gas properties from SINFONI data. The inferred results on ionised outflows are then presented in Sect. \ref{sec4}, whereas in Sect. \ref{sec5} they are studied as a function of host galaxy properties and compared with those previously obtained for the SUPER type-1 AGN \citep{Kakkad2020}. Further discussion on our main findings as well as comparison with results from the literature is found in Sect. \ref{sec6}. We finally draw our conclusions in Sect. \ref{sec7}. A $\Lambda$CDM flat cosmology with $\Omega_{\rm m,0} = 0.3$, $\Omega_{\rm \Lambda,0} = 0.7$ and $H_{\rm 0} = 70$ km s$^{-1}$ Mpc$^{-1}$ is adopted throughout this work.

\section{Sample and data description}\label{sec2}
The sample presented in this work is part of the SUPER survey (ID: 196.A-0377; PI: V. Mainieri), which acquired near-IR IFS observations of 33\footnote{Compared to the original 39-AGN SUPER sample, six objects were not observed eventually. This was a consequence of time lost because of bad weather conditions, since part of the survey was executed in visitor mode rather than in service.} blindly selected X-ray AGN at $z\sim2$ (see \citealt{Circosta2018} for details on sample selection and properties). The total 33-AGN observed sample consists of: 21 type-1 AGN, deeply investigated in \citet{Kakkad2020} and \citet{Kakkad2023}; and 12 type-2 AGN, whose SINFONI observations are presented and analysed in this work. All together these three papers investigate ionised outflows with an unprecedented spatial resolution of $\sim$ kpc in an unbiased sample of AGN at cosmic noon, using [O\,{\sc iii}] line emission as outflow tracer, and search for links between outflow properties (e.g. velocity, mass rate) and fundamental AGN/host galaxy parameters (e.g. bolometric luminosity, stellar mass, star formation rate). In the following subsections, we first introduce the observed type-2 AGN subsample (Sect. \ref{sec:super_ty2sample}), then we briefly describe the observations, the data reduction, and the final type-2 AGN sample (i.e. seven objects) analysed in this paper (Sect. \ref{sec:super_obs}).

\begin{table*}
\centering
\caption{Main properties of the 12 observed type-2 AGN from SUPER. From left, columns report: sky field and target ID; coordinates of the optical counterpart of the target (J2000); spectroscopic redshift $z$; galaxy stellar mass M$_{\ast}$, star formation rate (SFR), AGN bolometric luminosity L$_{\rm bol}$, all with 1$\sigma$ uncertainties, derived from SED fitting \citep{Circosta2018}; 2--10 keV X-ray luminosity $L_{\rm 2-10~keV}$ and hydrogen column density N$_{\rm H}$ with 90\% confidence level errors, measured from archival X-ray data \citep{Circosta2018}; whether the object is detected in SINFONI observations, hence included in the final analysed sample (see Sect. \ref{sec:super_obs}). The redshift of the detected objects is inferred from rest-frame optical SINFONI spectra (Sect. \ref{sec:super_specfit}), whereas for the others from archival rest-frame UV spectra (marked with $^{\dagger}$).}
\label{tab:super_ty2sample}
\resizebox{0.95\linewidth}{!}{
\begin{tabular}{l|cclccccc|c}
\hline
Field \& Target ID & RA & DEC & ~~~$z$ & log M$_{\ast}$ & SFR & log $L_{\rm bol}$ & log $L_{\rm 2-10~keV}$ & log $N_{\rm H}$ & Analysed\\
 & \footnotesize{(hh:mm:ss)} & \footnotesize{(dd:mm:ss)} & & \footnotesize{[M$_{\odot}$]} & \footnotesize{[M$_{\odot}$ yr$^{-1}$]} & \footnotesize{[erg s$^{-1}$]} & \footnotesize{[erg s$^{-1}$]} & \footnotesize{[cm$^{-2}$]} & \\
\hline
\\
CDF-S XID36 & 03:31:50.77 & $-$27:47:03.41 & 2.255 & $10.68 \pm 0.07$ & $184 \pm 9$ &  $45.70 \pm 0.06$ & $43.84^{+0.31}_{-0.63}$ &  $>24.1$ & yes\\

CDF-S XID57 & 03:31:54.40 & $-$27:56:49.70 & 2.298 $^{\dagger}$ & $10.49 \pm 0.11$ & $< 34$ & $44.26 \pm 0.18$ & $44.04^{+0.17}_{-0.24}$ &  $23.30^{+0.32}_{-0.39}$ & no\\

CDF-S XID419 &03:32:23.44 & $-$27:42:54.97 & 2.143  & $10.89 \pm 0.02$ & $42 \pm 4$ & $45.54 \pm 0.05$ & $43.84^{+0.29}_{-0.44}$ & $24.28^{+0.19}_{-0.31}$ & yes\\

CDF-S XID427 &03:32:24.20 & $-$27:42:57.51 & 2.303 $^{\dagger}$ & $10.87 \pm 0.08$ & $< 72$ & $44.60 \pm 0.13$ & $43.20^{+0.06}_{-0.06}$ & $22.43^{+0.24}_{-0.34}$ & no\\

CDF-S XID522 & 03:32:28.50 & $-$27:46:57.99 & 2.309 $^{\dagger}$ & $10.42 \pm 0.02$ & $492 \pm 25$ &  $45.02 \pm 0.02$ & $43.51^{+0.76}_{-0.87}$ &$>22.5$ & no\\

CDF-S XID614 & 03:32:33.02 & $-$27:42:00.33 & 2.453 & $10.78 \pm 0.08$ & $247 \pm 12$ & $44.97 \pm 0.13$ & $43.61^{+0.18}_{-0.18}$ & $24.25^{+0.19}_{-0.18}$ & yes\\

COSMOS cid\_1057 & 09:59:15.00 & +02:06:39.65 & 2.210 & $10.84 \pm 0.07$ & $85 \pm 4$ & $45.91 \pm 0.06$ & $44.53^{+0.26}_{-0.30}$ & $23.98^{+0.24}_{-0.28}$ & yes\\

COSMOS cid\_451 & 10:00:00.61 & +02:15:31.06 & 2.444 & $11.21 \pm 0.05$ & $<125$ & $46.44 \pm 0.07$ & $45.18^{+0.23}_{-0.19}$ & $23.87^{+0.19}_{-0.15}$ & yes\\

COSMOS cid\_2682 &10:00:08.81 & +02:06:37.66 & 2.437 & $11.03 \pm 0.04$ & $<93$ & $45.48 \pm 0.10$ & $44.30^{+0.96}_{-0.27}$ & $23.92^{+1.01}_{-0.20}$ & yes\\

COSMOS cid\_1143 &10:00:08.84 & +02:15:27.99 & 2.443 & $10.40 \pm 0.17$ & $108 \pm 18$ & $44.85 \pm 0.12$ & $44.83^{+0.43}_{-0.36}$ & $24.01^{+0.77}_{-0.29}$ & yes\\

COSMOS cid\_971 & 10:00:59.45 & +02:19:57.44 & 2.473 $^{\dagger}$ & $10.60 \pm 0.12$ & $<96$ & $44.71 \pm 0.24$ & $43.87^{+0.36}_{-0.38}$ & $<23.68$ & no\\

COSMOS cid\_1253 & 10:01:30.57 & +02:18:42.57 & 2.147 $^{\dagger}$ & $10.99 \pm 0.25$ & $280 \pm 194$ & $45.08 \pm 0.18$ & $43.92^{+0.29}_{-0.31}$ & $23.22^{+0.47}_{-0.39}$ & no\\
\\
\hline
\end{tabular}
}
\end{table*}


\subsection{Observed SUPER type-2 AGN subsample}\label{sec:super_ty2sample}

SUPER targeted 12 type-2 AGN in total: six sources are from the \textit{Chandra} Deep Field-South (CDF-S; \citealt{Luo2017}); whereas the other six are from the COSMOS-Legacy survey \citep{Civano2016}. Before being observed with SINFONI, they were identified as type-2 systems based on the absence of BLR emission in the archival rest-frame UV spectra \citep{Circosta2018}. Their type-2 spectral classification was then confirmed by SINFONI observations, showing no BLR components in rest-frame optical Balmer hydrogen lines (H$\alpha$ and H$\beta$), typically less extincted than UV emission lines. In Table \ref{tab:super_ty2sample}, we list some main properties of the 12 observed type-2 AGN, including spectroscopic redshift $z$; bolometric luminosity $L_{\rm bol}$, stellar mass $M_*$, and SFR, all derived from Spectral Energy Distribution (SED) fitting \citep{Circosta2018}; 2--10 keV X-ray luminosity $L_{\rm 2-10~keV}$ and hydrogen column density $N_{\rm H}$ inferred from archival X-ray spectra \citep{Circosta2018}. Redshift values have been obtained from rest-frame optical SINFONI spectra (see Sect. \ref{sec:super_specfit}), except for sources being undetected or marginally detected (five in total) in our SINFONI data (see Sect. \ref{sec:super_obs}). For these sources, the redshift has been measured from archival rest-frame UV spectra (marked with $^{\dagger}$). Depending on the quality of [O\,{\sc iii}] detection in SINFONI \textit{H}-band spectra, the last column shows what objects are included in the final type-2 AGN sample (i.e. seven objects) analysed in this work (details in Sect. \ref{sec:super_obs}).

\subsection{Observations and data reduction}\label{sec:super_obs}

\begin{table*}
\centering
\caption{SINFONI observations of the 12 SUPER type-2 AGN. For each target, we list the observing mode (AO/noAO) and the following parameters for each adopted grating (\textit{H}, \textit{K}, \textit{HK}): absolute (AB) magnitude measured from SINFONI integrated spectra accounting for the corresponding filter transmission curve, total on-source exposure time $t_{\rm exp}$, spatial resolution $\theta_{\rm PSF}$ (computed as PSF FWHM), and detected emission lines for the objects with {\normalfont [O\,{\sc iii}]} detected at S/N$>$2. For these sources, we indicate marginal detections (S/N$<$2) of {\normalfont H$\alpha$} with a ‘$^{\dagger}$' symbol. The AB magnitudes marked with ‘$^*$' have required a correction factor ($\sim 2$) to match the previous synthetic measurements \citep{Circosta2018}.}
\resizebox{0.95\linewidth}{!}{
\begin{tabular}{l|c|cccc|cccc}
\hline
 Target   & \footnotesize{Obs mode} & \multicolumn{4}{@{}c@{}}{\textit{H} band} & \multicolumn{4}{@{}c@{}}{\textit{K} band}\\
 & &  \multirow{2}{*}{mag (AB)} & \multirow{2}{*}{$t_{\rm exp}$ [hr]} & $\theta_{\rm PSF}$ [$''$] & \multirow{2}{*}{Detected lines} & \multirow{2}{*}{mag (AB)} & \multirow{2}{*}{$t_{\rm exp}$ [hr]} & $\theta_{\rm PSF}$ [$''$] & \multirow{2}{*}{Detected lines}\\
 & & & & ($\theta_{\rm PSF}$ [kpc]) & & & & ($\theta_{\rm PSF}$ [kpc]) & \\
\hline
\hline
\multirow{2}{*}{XID36} & \multirow{2}{*}{AO} & \multirow{2}{*}{21.45} & \multirow{2}{*}{5.3} & $0.29\times 0.26$ & \multirow{2}{*}{H$\beta$, [O\,{\sc iii}]} & \multirow{2}{*}{21.13} & \multirow{2}{*}{2.0} & $0.15\times0.15$ & \multirow{2}{*}{H$\alpha$, [N\,{\sc ii}], [S\,{\sc ii}]}\\
 & & & & ($2.4\times2.1$) & & & & $1.2\times1.2$ & \\
\hline
\multirow{2}{*}{XID57} & \multirow{2}{*}{AO} & \multirow{2}{*}{-} & \multirow{2}{*}{3.0} & $0.32\times 0.30$ & \multirow{2}{*}{-} & \multirow{2}{*}{-} & \multirow{2}{*}{1.0} & $0.33\times 0.27$ & \multirow{2}{*}{-}\\
 & & & & ($2.6\times2.5$) & & & & ($2.7\times2.2$) & \\
\hline
\multirow{2}{*}{XID419} & \multirow{2}{*}{AO} & \multirow{2}{*}{22.69*} & \multirow{2}{*}{0.8} & $0.47\times 0.52$ & \multirow{2}{*}{[O\,{\sc iii}]} & \multirow{2}{*}{21.85*} & \multirow{2}{*}{1.0} & $0.24\times 0.22$ & \multirow{2}{*}{H$\alpha$, [N\,{\sc ii}]}\\
 & & & & ($3.9\times4.3$) & & & & ($2.0\times1.8$) & \\
\hline
\multirow{2}{*}{XID427}  & \multirow{2}{*}{AO} & \multirow{2}{*}{22.72} & \multirow{2}{*}{1.0} & $0.23\times 0.20$ & \multirow{2}{*}{-} & \multirow{2}{*}{21.90*} & \multirow{2}{*}{1.0} & $0.28\times0.20$ & \multirow{2}{*}{-}\\
 & & & & ($1.9\times1.6$) & & & & ($2.3\times1.6$) & \\
\hline
\multirow{2}{*}{XID522}  & \multirow{2}{*}{AO} & \multirow{2}{*}{-} & \multirow{2}{*}{1.0} & $0.25\times0.20$ & \multirow{2}{*}{-} & \multirow{2}{*}{-} & \multirow{2}{*}{1.0} & $0.25\times0.19$ & \multirow{2}{*}{-}\\
 & & & & ($2.0\times1.6$) & & & & ($2.0\times1.6$) & \\
\hline
\multirow{2}{*}{XID614} & \multirow{2}{*}{AO} & \multirow{2}{*}{22.41*} & \multirow{2}{*}{11.5} & $0.38\times 0.37$ & \multirow{2}{*}{H$\beta$, [O\,{\sc iii}]} & \multirow{2}{*}{21.79} & \multirow{2}{*}{3.5} & $0.27\times 0.28$ & \multirow{2}{*}{H$\alpha$, [N\,{\sc ii}]}\\
 & & & & ($3.1\times3.0$) & & & & ($2.2\times2.3$) & \\
\hline
\multirow{2}{*}{cid\_1057} & \multirow{2}{*}{AO} & \multirow{2}{*}{21.80*} & \multirow{2}{*}{2.0} & $0.32\times 0.30$ & \multirow{2}{*}{H$\beta$, [O\,{\sc iii}]} & \multirow{2}{*}{20.91} & \multirow{2}{*}{0.7} & $0.46\times 0.43$ & \multirow{2}{*}{-}\\
 & & & & ($2.6\times2.5$) & & & & ($3.8\times3.6$) & \\
\hline
\multirow{2}{*}{cid\_451} & \multirow{2}{*}{AO} & \multirow{2}{*}{21.73} & \multirow{2}{*}{5.0} & $0.28\times 0.30$ & \multirow{2}{*}{[O\,{\sc iii}]} & \multirow{2}{*}{21.69*}  & \multirow{2}{*}{2.0} & $0.28\times 0.27$ & \multirow{2}{*}{H$\alpha$, [N\,{\sc ii}]}\\
 & & & & ($2.3\times2.4$) & & & & ($2.3\times2.2$) & \\
\hline
\multirow{2}{*}{cid\_2682} & \multirow{2}{*}{AO} & \multirow{2}{*}{21.58} & \multirow{2}{*}{5.6} & $0.33\times0.28$ & \multirow{2}{*}{[O\,{\sc iii}]} & \multirow{2}{*}{21.01*} & \multirow{2}{*}{1.8} & $0.28\times 0.25$ & \multirow{2}{*}{H$\alpha$, [N\,{\sc ii}]}\\
 & & & & ($2.7\times2.3$) & & & & ($2.3\times2.0$) & \\
\hline
\multirow{2}{*}{cid\_1143} & \multirow{2}{*}{AO} & \multirow{2}{*}{22.44} & \multirow{2}{*}{5.5} & $0.41\times 0.38$ & \multirow{2}{*}{H$\beta$, [O\,{\sc iii}]} & \multirow{2}{*}{22.66}  & \multirow{2}{*}{2.0} & $0.30\times 0.30$ & \multirow{2}{*}{H$\alpha$ $^{\dagger}$}\\
 & & & & ($3.3\times3.1$) & & & & ($2.4\times2.4$) & \\
\hline
\multirow{2}{*}{cid\_1253} & \multirow{2}{*}{noAO} & \multirow{2}{*}{22.45} & \multirow{2}{*}{4.0} & $0.68\times 0.63$ & \multirow{2}{*}{-} & \multirow{2}{*}{21.61} & \multirow{2}{*}{1.0} & $0.66\times 0.61$ & \multirow{2}{*}{-}\\
 & & & & ($5.6\times5.2$) & & & & ($5.5\times5.1$) & \multirow{2}{*}{ }\\
\hline
\hline
& & \multicolumn{8}{@{}c@{}}{\textit{HK} band}\\
& & \multicolumn{2}{@{}c@{}}{mag (AB)} & \multicolumn{2}{@{}c@{}}{$t_{\rm exp}$ [hr]} & \multicolumn{2}{@{}c@{}}{$\theta_{\rm PSF}$ [$''$]} & \multicolumn{2}{@{}c@{}}{Detected lines}\\
& & \multicolumn{2}{@{}c@{}}{} & \multicolumn{2}{@{}c@{}}{} & \multicolumn{2}{@{}c@{}}{($\theta_{\rm PSF}$ [kpc])} & \multicolumn{2}{@{}c@{}}{}\\
\hline
\multirow{2}{*}{cid\_971}  & \multirow{2}{*}{noAO} & \multicolumn{2}{c}{\multirow{2}{*}{-}} & \multicolumn{2}{c}{\multirow{2}{*}{1.8}} & \multicolumn{2}{c}{$0.42\times 0.39$} & \multicolumn{2}{c}{\multirow{2}{*}{-}}\\
 & & \multicolumn{2}{c}{} & \multicolumn{2}{c}{} & \multicolumn{2}{c}{($3.4\times 3.2$)} & \multicolumn{2}{c}{}\\
\hline
\end{tabular}
}
\label{tab:super_obs}
\end{table*}

SUPER targets were observed with SINFONI between November 2015 and December 2018.
All observations were planned to be carried out in AO mode 
using a laser guide star (LGS), covering
both \textit{H} band, targeting H$\beta$ and [O\,{\sc iii}]$\lambda\lambda$4959,5007 emission lines, and \textit{K} band, targeting H$\alpha$, [N\,{\sc ii}]$\lambda\lambda$6549,83, and [S\,{\sc ii}]$\lambda\lambda$6716,31.

Due to the absence of suitable stars close to the targets, observations of ten out of 12 type-2 AGN were performed in AO-assisted mode with no tip-tilt star (i.e. in Seeing Enhancer mode; \citealt{Davies2008}), using both \textit{H} and \textit{K} gratings ($R\sim$3000 and $R\sim$4000, respectively); the remaining two objects were observed with no-AO (cid\_1253 with \textit{H} and \textit{K} gratings; cid\_971 with the \textit{HK} grating, $R\sim$1500) in order to optimize the time available during the visitor-mode observing run, reduced because of bad weather conditions. For AO-assisted observations, we adopted a plate scale of $3''\times 3''$ with a spatial sampling of $0.05''\times 0.1''$, then re-sampled to $0.05''\times 0.05''$ in the final datacube. For the two targets observed in seeing-limited mode, we instead selected the largest field of view (FoV; $8''\times 8''$), corresponding to a pixel scale of $0.25''\times 0.25''$. Before or after each observing block, a dedicated bright star was observed to estimate the Point Spread Function (PSF) of our AO-assisted SINFONI observations, as full width at half maximum (FWHM) of a 2D-Gaussian fitting to the total flux distribution of the PSF star ($\theta_{\rm PSF}$ values in Table \ref{tab:super_obs}).

Below we summarise the main steps of the reduction procedure applied to SINFONI type-2 AGN datacubes, whereas we refer the reader to \citet{Kakkad2020} for a detailed description of the SUPER observational strategy. We reduced all observations using the ESOREX pipeline (3.1.1), which returns a distortion-corrected and wavelength-calibrated datacube of the science target, as well as of the PSF and telluric stars. We then removed background sky emission via the \texttt{IDL} routine ‘skysub.pro' \citep{Davies2007}, and used our own custom-made \texttt{python} routines to perform flux calibration (based on \citealt{Piqueras2012}) and to reconstruct the final datacube for each observed target. As a final check, we compared the synthetic photometry measured from SINFONI integrated spectra with both archival photometric measurements \citep{Circosta2018}, and the synthetic photometry from KMOS integrated spectra, when available (Scholtz et al., in prep.). From this comparison, we estimated a typical relative uncertainty on the flux calibration of our SINFONI data of about 20\% for all employed bands.

In Table \ref{tab:super_obs}, we show the main parameters of SINFONI observations of the 12 SUPER type-2 AGN. For each target, we specify the observing mode (AO/noAO), and list the following parameters for the corresponding adopted grating (\textit{H}, \textit{K}, \textit{HK}): absolute (AB) magnitude measured from SINFONI integrated spectra, total exposure time $t_{\rm exp}$, spatial resolution $\theta_{\rm PSF}$ (i.e. PSF FWHM), and detected emission lines for the seven objects with [O\,{\sc iii}] detected with a total, spatially integrated S/N$>$2, that is the crucial tracer of ionised outflows employed in our study. Five of these systems have also H$\alpha$+[N\,{\sc ii}] detected with a S/N$>$2, whereas H$\alpha$ is totally undetected or marginally detected (S/N$<$2) in cid\_1057 and cid\_1143, respectively. Although not suitable for our analysis, we report for completeness that we also marginally detect [O\,{\sc iii}] and/or H$\alpha$ in XID427 and cid\_1253. Finally, the remaining three targets (i.e. XID522, XID57 and cid\_971) are totally undetected in SINFONI  observations. We point out that these non-detections or poor detections are a consequence of the lower integration time, due to the observing time lost because of bad weather conditions, which particularly penalised SUPER type-2 objects, being these also the faintest sources of the SUPER sample ($L_{\rm bol} < 10^{46}$ erg s$^{-1}$).

In the following, this paper will focus on the seven type-2 AGN with [O\,{\sc iii}] detected at S/N$>$2, which allows us to perform a spatially-resolved analysis of [O\,{\sc iii}] line emission and thus search for ionised outflows signatures. These sources have all a ‘yes' associated in the last column of Table \ref{tab:super_ty2sample}. In particular, three of them are CDF-S sources (XID36, XID419, XID614), whereas four are from COSMOS (cid\_1057, cid\_1143, cid\_2682, cid\_451).

\section{Data analysis}\label{sec3}
In this section, we analyse SINFONI \textit{H}-band observations of the seven type-2 AGN well detected (S/N$>$2) in [O\,{\sc iii}], to spatially map [O\,{\sc iii}] line emission and its kinematics, looking for evidence of high-velocity outflowing gas (Sect. \ref{sec:super_specfit}). Unfortunately, the quality of \textit{K}-band data is worse (i.e. lower S/N) for most objects, thus preventing us from an accurate mapping of \textit{K}-band line emission across the SINFONI FoV. 
In Sect. \ref{sec:super_integrated} we use combined integrated measurements of H$\alpha$ and H$\beta$ fluxes from \textit{K}- and \textit{H}-band data, respectively, to estimate dust extinction. Being these targets obscured (type-2) AGN, correcting for dust extinction is a crucial step to get reliable estimates of outflow energetics (see Sect. \ref{sec:super_energetics}).

\subsection{Spectral fitting}\label{sec:super_specfit}

\begin{figure*}
    \centering
    \includegraphics[width=0.98\linewidth]{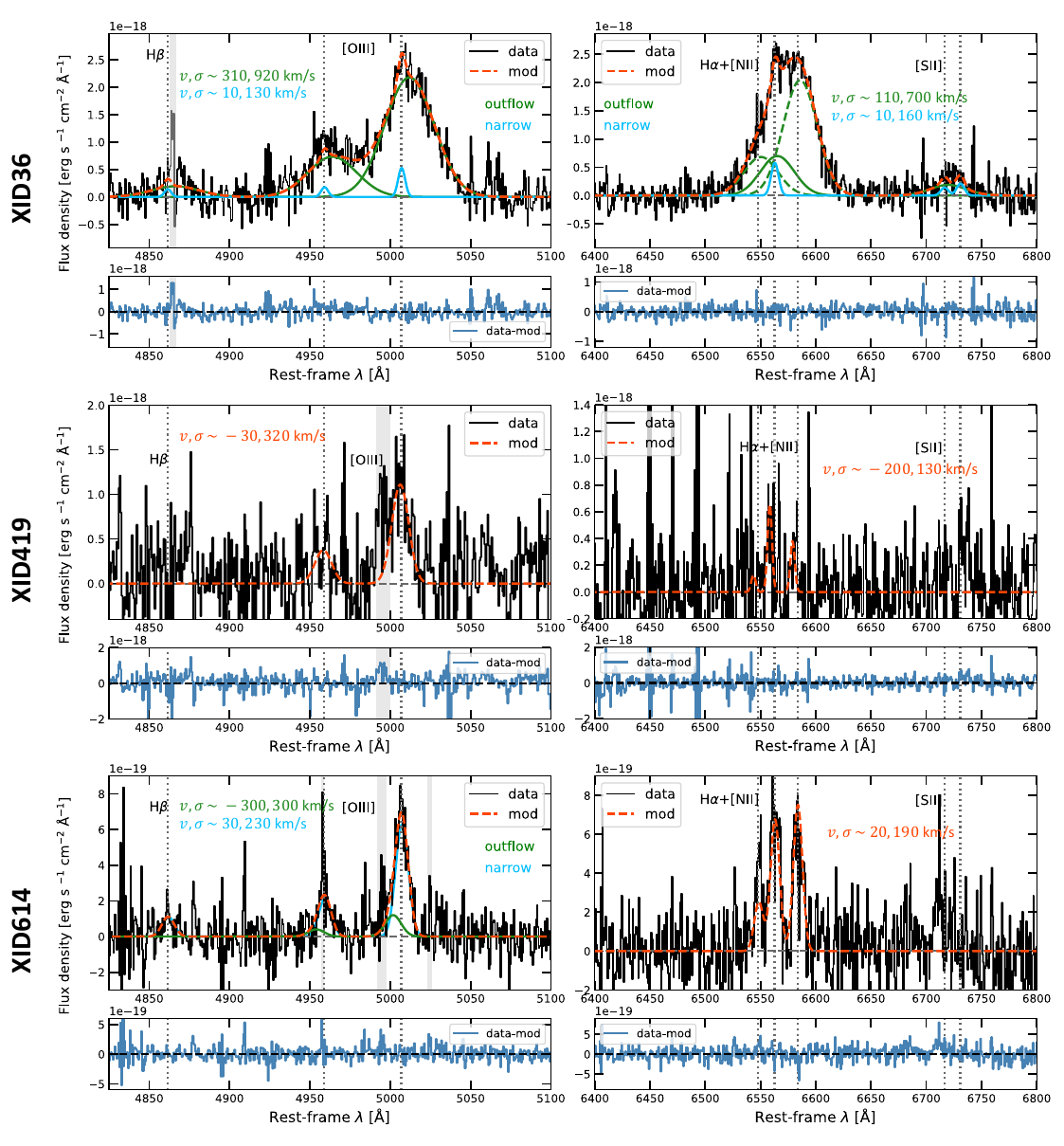}
    \caption{Integrated \textit{H}- and \textit{K}-band spectra extracted from SINFONI datacubes of XID36 (top), XID419 (middle) and XID614 (bottom), with a 0.25$''$-radius aperture, after subtracting continuum emission modelled with a 1st-degree polynomial. Data are shown in black, with total emission-line model drawn on the top in red. For multi-Gaussian modellings, we plot separately narrow, systemic components (light blue) and broad high-velocity ones (green), associated with outflows. In the top right panel, green solid and dashed lines represent the outflow component of H$\alpha$ and [N\,{\sc ii}] doublet, respectively. Mean velocity $v$ and velocity dispersion $\sigma$ values of each Gaussian component are displayed. Shaded grey regions indicate masked channels contaminated by sky line residuals, while vertical dotted lines mark rest-frame emission line wavelengths at the redshift of each source, as computed in Sect. \ref{sec:super_specfit}. Below each main panel, a second one shows corresponding residuals (i.e. data--model).}
    \label{fig:super_spectra_XID}
\end{figure*}

\begin{figure*}
    \centering
    \includegraphics[width=0.98\textwidth]{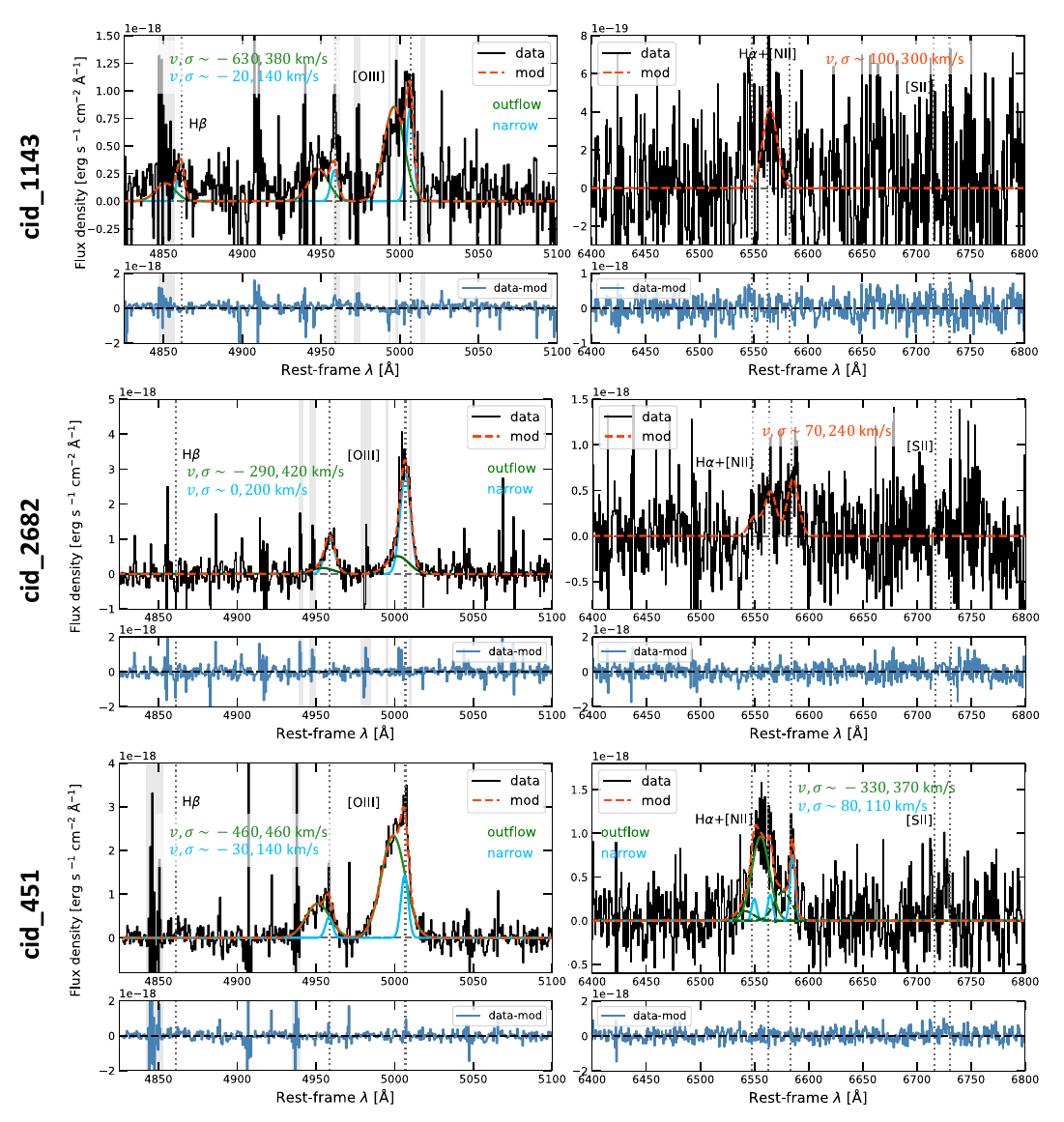}
    \caption{Integrated \textit{H}- and \textit{K}-band spectra extracted from SINFONI datacubes of cid\_1143 (top), cid\_2682 (middle) and cid\_451 (bottom), with a 0.25$''$-radius aperture, after subtracting continuum emission modelled with a 1st-degree polynomial. Same as in Fig. \ref{fig:super_spectra_CID}. In \textit{K}-band spectrum of cid\_1143, we fit only {\normalfont H$\alpha$} line emission due to the low S/N (S/N$<$2) of this dataset.}
    \label{fig:super_spectra_CID}
\end{figure*}

\begin{figure}
    \centering
    \includegraphics[width=0.98\columnwidth]{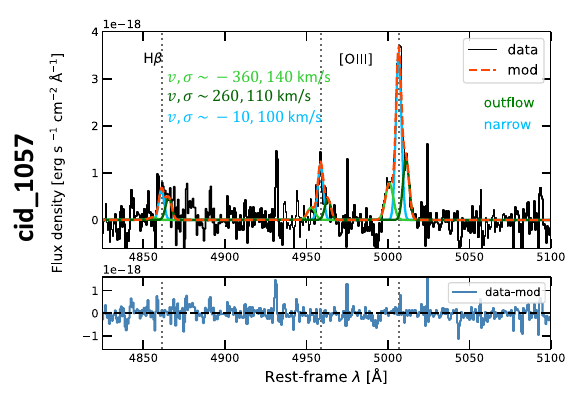}
    \caption{Integrated \textit{H}-band spectrum of cid\_1057, extracted from SINFONI datacube with a 0.25$''$-radius aperture, after subtracting continuum emission modelled with a 1st-degree polynomial. Same as in Fig. \ref{fig:super_spectra_CID}. Three Gaussian components are used to properly model {\normalfont [O\,{\sc iii}]} line profile: one for systemic, narrow line emission (light blue), two for the blue and red wings detected in {\normalfont [O\,{\sc iii}]} (different green colours).}
    \label{fig:super_spectra_CID1057}
\end{figure}

For the spectral fitting of SINFONI \textit{H}-band datacubes, we adopt the fitting code presented in \citet{Marasco2020} (see also \citealt{Tozzi2021} for more details), and fit all cubes in the rest-frame optical wavelength range of 4800--5200\AA. Being distant type-2 AGN, their SINFONI spectra show neither BLR emission nor strong (AGN and/or stellar) continuum, but consist of only ‘narrow' (FWHM$\lesssim$2000 km s$^{-1}$) emission lines, either originating from AGN Narrow Line Region (NLR) or due to a different kinematics (i.e. outflow, disk rotation). In addition to [O\,{\sc iii}]$\lambda\lambda$5007,4959 emission line doublet (detected in all seven examined sources), we also detect faint H$\beta$ line emission in four \textit{H}-band datacubes (i.e. XID36, XID914, cid\_1143, cid\_1057; see Figs. \ref{fig:super_spectra_XID}-\ref{fig:super_spectra_CID1057}).

As a first step, we apply a spatial Gaussian smoothing of the data channel by channel to enhance the visibility of real gas structures. To do this without deteriorating the instrumental PSF, we smooth each datacube with a Gaussian kernel of width $\sigma_{\rm smooth}\sim0.1-0.2''$, that is as large as possible, yet still smaller than the instrumental PSF FWHM (i.e. $\sigma_{\rm smooth}<\theta_{\rm PSF}/2.35$). After this, we model spaxel-by-spaxel the observed line emission via multiple Gaussian components, following the prescriptions summarised below:
\begin{itemize}
\item We fit the [O\,{\sc iii}]$\lambda\lambda$5007,4959 line doublet with two Gaussian components sharing the same kinematics (mean velocity $v$ and velocity dispersion $\sigma$), imposing a fixed flux ratio of 3. Hereafter, we refer to the brighter doublet component (i.e. [O\,{\sc iii}]$\lambda$5007) simply as [O\,{\sc iii}].
\item When detected, faint H$\beta$ line emission is modelled with $v$ and $\sigma$ fixed to the best-fit values obtained for [O\,{\sc iii}], hence with flux as the only free parameter.
\item In each spaxel, we reiterate the line modelling using an increasing number of Gaussian components for each emission line, from one to the maximum number which best reproduces complex line profiles in high S/N spaxels. We then select the optimal (minimum) number of Gaussian components required in each spaxel via a Kolmogorov-Smirnov test on the residuals. Five \textit{H}-band datasets require up to two Gaussian components; cid\_1057 up to three components, given the presence of both blue and red [O\,{\sc iii}] wings detected; instead, one component is sufficient to model single-spaxel line emission in XID419.
\item We also add a 1st-degree polynomial to reproduce any faint continuum emission over the entire fitted wavelength range.
\end{itemize}

Several spectra require multiple Gaussian components to reproduce simultaneously both narrow, low-velocity line emission and broad, high-velocity wings in [O\,{\sc iii}] line profile. Whereas the latter are considered the typical signature of fast outflowing gas (discussed in Sect. \ref{sec:super_kinematics}), narrow, low-velocity line emission is typically due to nearly systemic gas motions (e.g. NLR, disk rotation). In fact, we use the narrowest [O\,{\sc iii}] line component to measure the redshift of each source: we extract a \textit{H}-band integrated spectrum from a small aperture (0.1$''$-radius) centred on the overall \textit{H}-band emission peak (corresponding to [O\,{\sc iii}] emission peak), and set to 0 km s$^{-1}$ the peak of the narrowest [O,\,{\sc iii}] line component. We estimate an uncertainty of $\Delta~z=0.001$ on our [O\,{\sc iii}]-based measurements of $z$, due to SINFONI \textit{H}-band spectral resolution ($R\sim$3000) and the typical S/N of the line.

\subsection{Dust extinction from integrated spectra}\label{sec:super_integrated}

Since the targets examined in this work are all type-2 AGN highly obscured in the X-rays ($N_{\rm H}\gtrsim10^{23}$cm$^{-2}$; \citealt{Circosta2018}), rest-frame optical emission is expected to be significantly affected by extinction effects. In particular, in Sect. \ref{sec:super_energetics} we correct our measurements of [O\,{\sc iii}] outflow luminosity for dust extinction to estimate intrinsic energetics of ionised outflows. Below, we derive rest-frame optical dust extinction from spatially integrated H$\alpha$/H$\beta$ ratios for the seven examined type-2 AGN.

To accurately correct for dust extinction, in principle we should: (i) account for spatial variations of H$\alpha$/H$\beta$ ratios across the FoV using spaxel-by-spaxel measurements; (ii) consider extinction effects exclusively on the outflow component, which may differ from those affecting the bulk of the ionised gas or any other kinematic component (e.g. disk rotation, AGN NLR). Yet, the S/N on H$\alpha$ and H$\beta$ in our SINFONI data is not high enough for such an accurate, spatial mapping of dust extinction. Indeed, their detection is mostly limited to central brighter spaxels, with H$\beta$ undetected in three sources (even in integrated spectra), whereas H$\alpha$ is totally undetected in cid\_1057 and marginally (S/N$<$2) detected in cid\_1143. Therefore, we extract integrated \textit{K}- and \textit{H}-band spectra of each object to increase the S/N on H$\alpha$ and H$\beta$, so as to estimate (or put constraints on) dust extinction. The only target of the 7-AGN sample for which this is not possible is cid\_1057, being totally undetected in H$\alpha$. Therefore, in the following we describe first how we estimate
(or constrain) dust extinction from SINFONI integrated spectra in the six objects with H$\alpha$ detected, and then assumptions adopted for cid\_1057.

We extract integrated \textit{K}- and \textit{H}-band spectra of each target using an aperture of $\sim$0.25$''$-radius centred on the source (Figs. \ref{fig:super_spectra_XID}, \ref{fig:super_spectra_CID}, \ref{fig:super_spectra_CID1057}), and fit them following the same prescriptions described in Sect. \ref{sec:super_specfit}.
In particular, Gaussian profiles fitted to H$\alpha$ and [N\,{\sc ii}]$\lambda\lambda$6549,83 are constrained to have the same velocity and velocity dispersion, with the additional constraint of a fixed flux ratio of 3 on the two [N\,{\sc ii}] doublet components.
In XID36, we model also faint [S\,{\sc ii}]$\lambda\lambda$6716,31 line emission, following the same prescriptions adopted for H$\beta$ in \textit{H}-band datasets. In most cases, one Gaussian component is enough to reproduce \textit{K}-band emission lines; XID36 and cid\_451 instead require an additional broader component ($\sigma\sim700$ km s$^{-1}$ in XID36, $\sigma\sim400$ km s$^{-1}$ in cid\_451).

In Figs. \ref{fig:super_spectra_XID} and \ref{fig:super_spectra_CID}, we show \textit{H}- and \textit{K}-band spectra extracted with a 0.25$''$-radius aperture from SINFONI datacubes, after subtracting the 1st-degree polynomial used to reproduce any residual faint continuum emission. Data are shown in black, with total emission-line model drawn on the top in red. For multi-Gaussian fittings, we separately plot components used to reproduce narrow, systemic line emission (light blue), and broad high-velocity line emission (green), likely associated with ionised outflows. For all spectra, we also display velocity $v$ and velocity dispersion $\sigma$
values of each Gaussian component, using same colours as for the corresponding Gaussian component. In Fig. \ref{fig:super_spectra_CID1057}, we also show an integrated \textit{H}-band spectrum of cid\_1057, extracted with the same 0.25$''$-radius aperture. In this source, the [O\,{\sc iii}] emission line exhibits an asymmetric profile which is well reproduced by three Gaussian components: one for systemic, narrow line emission (light blue), two for the blue and red [O\,{\sc iii}] wings (in different green colours). Excluding XID419, the only target requiring a single-Gaussian modelling, we find mean values of $\langle |v| \rangle\sim380$ km s$^{-1}$ and $\langle\sigma \rangle\sim440$ km s$^{-1}$ for the broad outflow (green) component, whereas $\langle |v| \rangle\sim20$ km s$^{-1}$ and $\langle\sigma \rangle\sim160$ km s$^{-1}$ for the narrow, nearly systemic one.

Since we do not detect broad, high-velocity wings in H$\beta$ line profile, possibly associated with outflows, in any object (in H$\alpha$ only in XID36 and cid\_451), we take integrated total H$\alpha$/H$\beta$ ratios to get an estimate of total dust extinction \citep{Calzetti2000}, and assume it to also affect outflow emission. For XID36, XID614 and cid\_1143, with both hydrogen lines detected, we estimate $A_V$ and relative uncertainty via error propagation. For cid\_1143, due to the low S/N on H$\alpha$ and H$\beta$, we obtain a value of $A_V$ consistent with 0 (i.e. no extinction) within the uncertainty ($A_V = 1.2 \pm 1.8$). Since $A_V$ must be non-negative, we adopt $A_V = 1.2^{+1.8}_{-1.2}$.
In XID419, cid\_2682 and cid\_451 instead, where we do not detect H$\beta$, we estimate a lower limit to $A_V$, considering a 2$\sigma_{\rm noise}$ upper limit to H$\beta$ flux. Finally, for cid\_1057, the only source totally undetected in H$\alpha$, we consider an $A_V$ value computed as the average of the three $A_V$ measurements obtained for XID36, XID614 and cid\_1143, taking the maximum distance of these estimates from the mean as uncertainty (i.e. $A_V=1.8\pm1.0$). We will use these $A_V$ values later in Sect. \ref{sec:super_energetics} to get the dust-corrected [O\,{\sc iii}] outflow flux in each source, hence the corresponding outflow mass.

\section{Results}\label{sec4}
\subsection{Evidence of spatially resolved outflows}\label{sec:super_kinematics}

\begin{figure*}
    \centering    \includegraphics[width=0.9\linewidth]{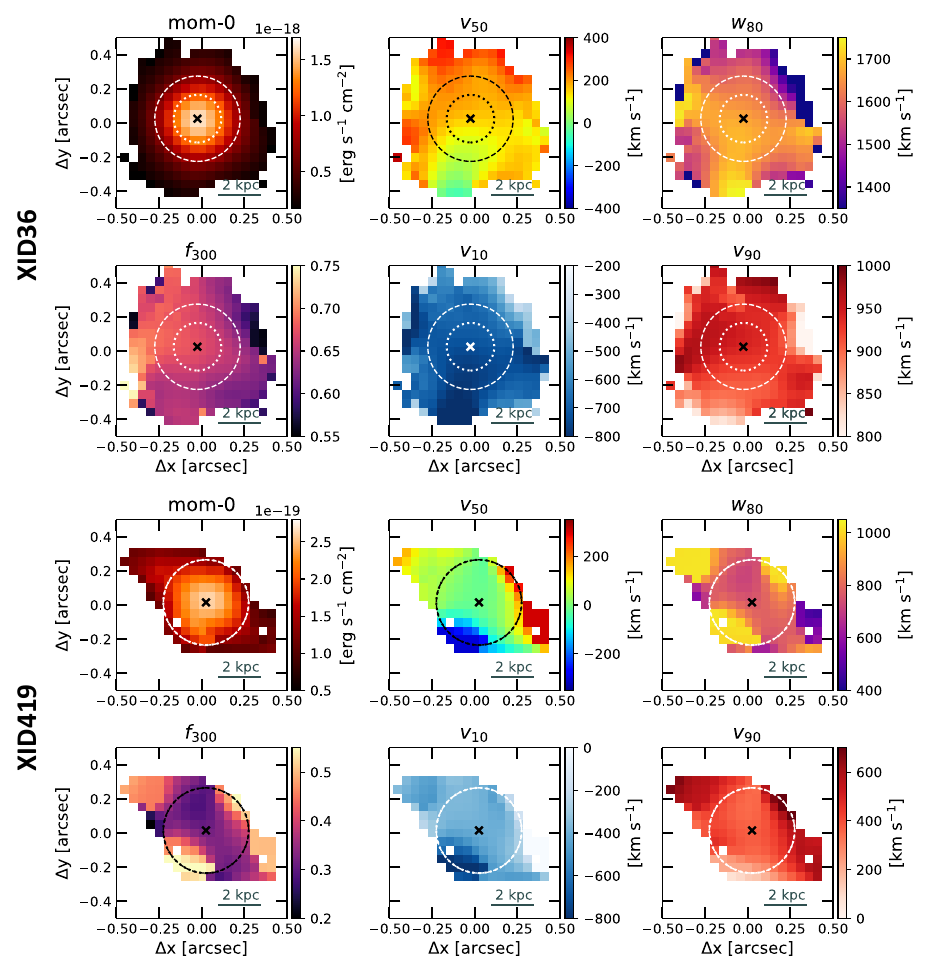}
    \caption{Morphology and kinematics of ionised gas in XID36 and XID419, as traced by total {\normalfont [O\,{\sc iii}]} line emission. For each target, maps show: the intensity (moment-0) field; $v_{\rm 50}$, $v_{\rm 10}$, $v_{\rm 90}$ percentile velocities and $w_{\rm 80}$ line width; and the ratio $f_{300}$ of the flux contained in {\normalfont [O\,{\sc iii}]} wings (i.e. $|v|>300$ km s$^{-1}$) to the moment-0 flux. Dashed and dotted circles correspond, respectively, to the 0.25$''$-radius aperture used to extract integrated spectra (Sect. \ref{sec:super_integrated}), and to the mean \textit{H}-band PSF of radius $\langle \theta_{\rm PSF}\rangle/2$ (see Table \ref{tab:super_obs}). In XID419 the two circles have the same 0.25$''$ radius. In all maps, we apply a S/N$>$3 cut and mark the position of {\normalfont [O\,{\sc iii}]} peak emission with a cross.}
    \label{fig:super_maps_XID}
\end{figure*}

\begin{figure*}
    \centering    \includegraphics[width=0.9\linewidth]{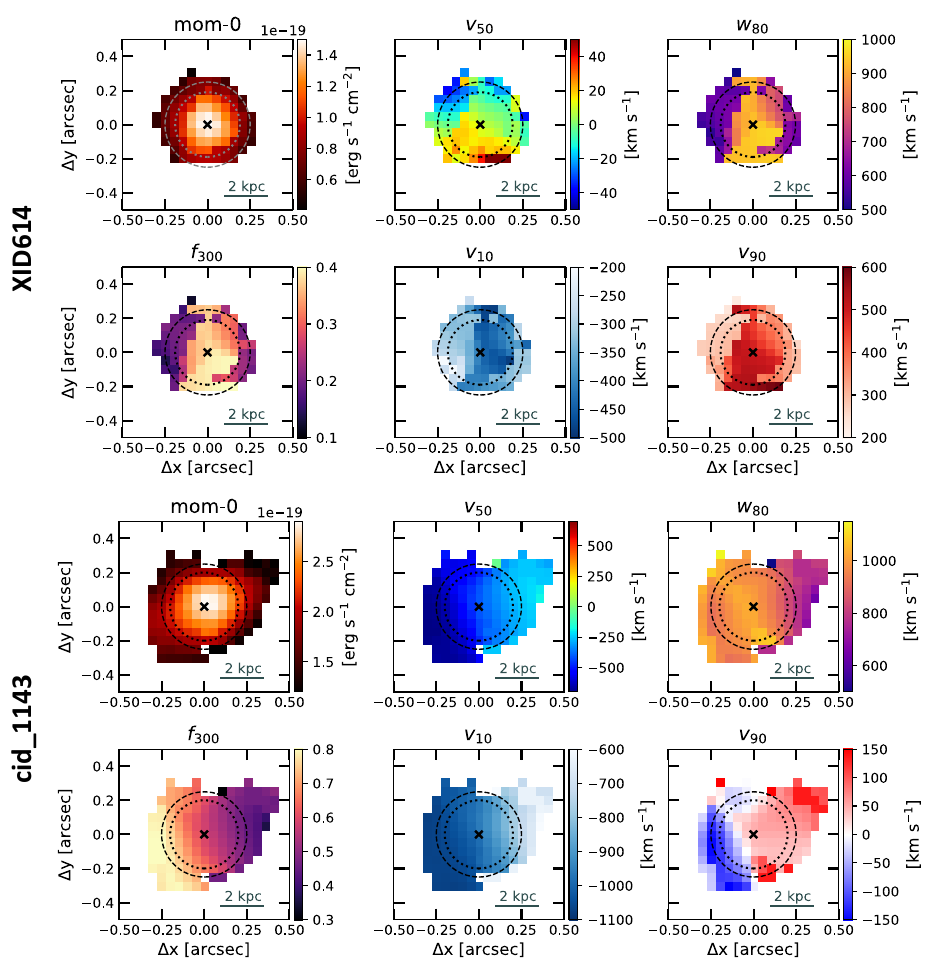}
    \caption{Morphology and kinematics of ionised gas in XID614 and cid\_1143, as traced by total {\normalfont [O\,{\sc iii}]} line emission. Same as in Fig. \ref{fig:super_maps_XID}.}
    \label{fig:super_maps_2}
\end{figure*}

\begin{figure*}
    \centering    \includegraphics[width=0.9\linewidth]{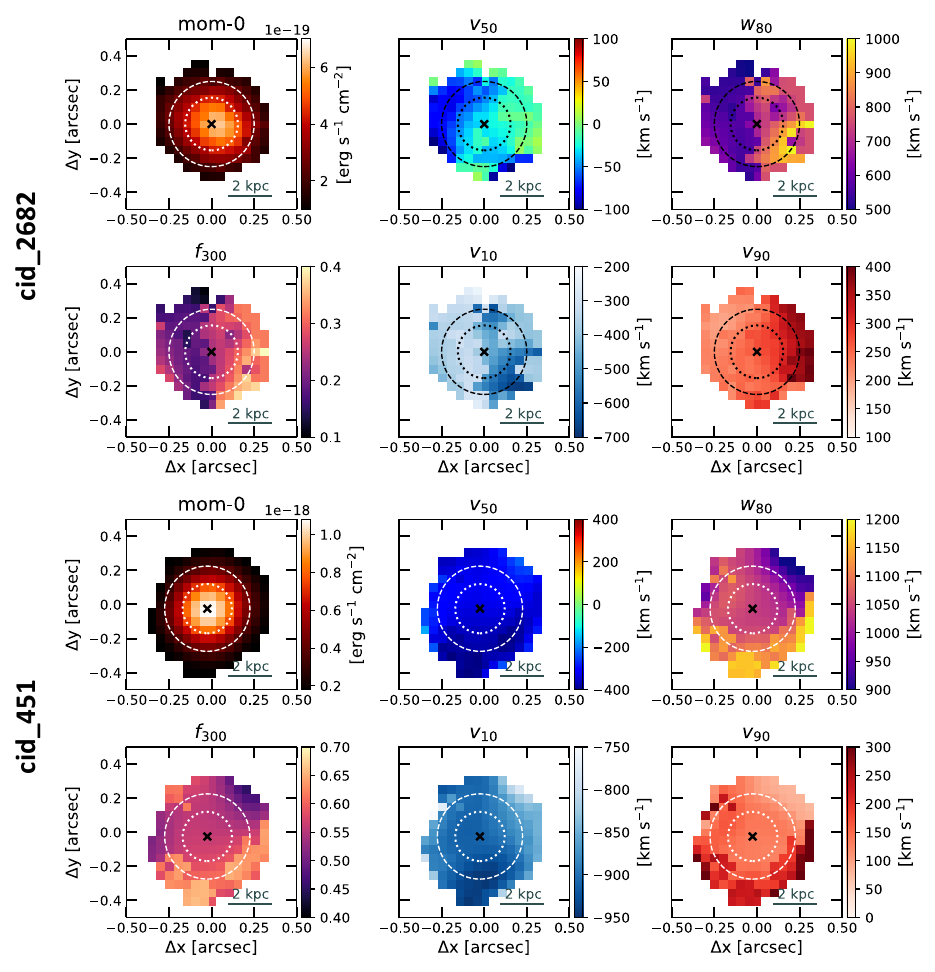}
    \caption{Morphology and kinematics of ionised gas in cid\_2682 and cid\_451, as traced by total {\normalfont [O\,{\sc iii}]} line emission. Same as in Fig. \ref{fig:super_maps_XID}.}
    \label{fig:super_maps_3}
\end{figure*}


\begin{figure*}
    \centering    \includegraphics[width=0.9\linewidth]{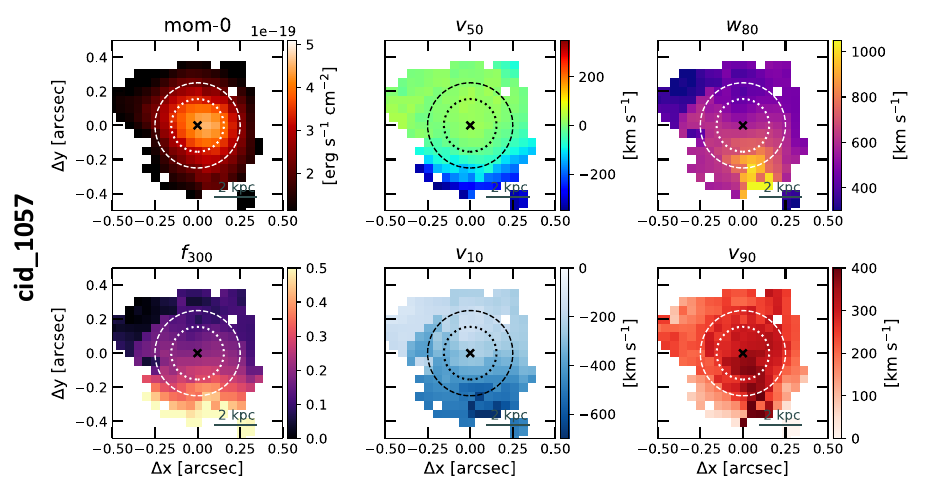}
    \caption{Morphology and kinematics of ionised gas in cid\_1057, as traced by total {\normalfont [O\,{\sc iii}]} line emission. Same as in Fig. \ref{fig:super_maps_XID}.}
    \label{fig:super_maps_CID1057}
\end{figure*}

Our spectral modelling of [O\,{\sc iii}] line emission has clearly revealed prominent wings in the [O\,{\sc iii}] line profile of four galaxies (i.e. XID36, cid\_1143, cid\_451, cid\_1057; see Figs. \ref{fig:super_spectra_XID}-\ref{fig:super_spectra_CID1057}), pointing to the presence of gas moving at high speed with respect to the galaxy systemic velocity. Also in XID614 and cid\_2682, the [O\,{\sc iii}] line profile shows a faint blue wing (see Figs. \ref{fig:super_spectra_XID}, \ref{fig:super_spectra_CID}), whereas it is hard to assess at first sight the presence of any possible blue [O\,{\sc iii}] wing in XID419 (see Fig. \ref{fig:super_spectra_XID}), due to bright residual sky emission. In this section we inspect kinematic maps of [O\,{\sc iii}] line emission (Figs. \ref{fig:super_maps_XID}-\ref{fig:super_maps_CID1057}) to confirm the presence of ionised outflows in higher-S/N detections, and to search for evidence in more ambiguous cases.

In Figs. \ref{fig:super_maps_XID}--\ref{fig:super_maps_CID1057}, we show maps of [O\,{\sc iii}] line emission with a S/N$>$3 cut, displaying the spatial distribution and kinematics of ionised gas in all seven examined type-2 AGN. First-column maps show the [O\,{\sc iii}] intensity field computed as moment-0 of the total line profile (upper panel); and ratio values ($f_{300}$; lower panel) of the flux in high-velocity channels ($|v|>300$ km s$^{-1}$) to moment-0 flux, a simplified but reasonable approximation of [O\,{\sc iii}] flux carried by outflows \citep{Kakkad2020}. For the seven examined type-2 AGN, we find that the total (i.e. spatially integrated) flux at $|v|>300$ km s$^{-1}$ represents a significant fraction of the total moment-0 flux, ranging from 21\% (in cid\_1057) to 66\% (in XID36).

In addition, maps in the second and third columns of Figs. \ref{fig:super_maps_XID}--\ref{fig:super_maps_CID1057} display ionised gas kinematics described in terms of non-parametric percentile velocities\footnote{Percentile velocities $v_{\rm xx}$ used in the following are defined as the velocity containing xx\% of the total [O\,{\sc iii}] line flux.}. Compared to parametric values, percentile velocities have the advantage of being independent on the adopted fitting function (e.g. the number of Gaussian components), which in turn may strongly depend on the S/N of the examined data (e.g. \citealt{Harrison2014,Zakamska2014}). In particular, we compute $v_{\rm 50}$, $v_{\rm 10}$, $v_{\rm 90}$ velocities, and the $w_{\rm 80}$ line width (i.e. $w_{\rm 80}=v_{\rm 90}-v_{\rm 10}$), approximately equal to the FWHM of a Gaussian profile. Extreme $v_{\rm 10}$ and $v_{\rm 90}$ values are widely adopted as a reliable approximation of outflow velocity, for its approaching and receding components (e.g. \citealt{Harrison2014,Carniani2015,Cresci2015}), respectively. Compared to moment-1 and moment-2 values (i.e. mean velocity and velocity dispersion, respectively), which may be affected by significant geometrical projection effects, $v_{\rm 10}$ and $v_{\rm 90}$ help avoid possible underestimates of the real outflow velocity, due to our ignorance about intrinsic outflow geometry and inclination to the line of sight. Maps of $v_{\rm 10}$ and $v_{\rm 90}$ overall feature high velocities, with absolute values up to 800--1100 km s$^{-1}$ in four targets (XID36, XID419, cid\_1143, cid\_451), and to 600--700 km s$^{-1}$ in the remaining three (XID614, cid\_1057, cid\_2682). These large velocities are accompanied by $w_{\rm 80}$ line widths larger than 800 km s$^{-1}$ in most of the FoV of all targets, with maximum values of at least 1000 km s$^{-1}$ (about 1700 km s$^{-1}$ in XID36 and 1200 km s$^{-1}$ in cid\_451). 

Ionised outflows are the most likely explanation to the high-velocity kinematics observed in all targets, since alternative phenomena can hardly lead to such high values of velocity and line width. In fact, SF-driven outflows are typically characterised by lower velocities (e.g. \citealt{Arribas2014,Foerster2019}); whereas galactic inflows are mostly observed in absorption with low bulk velocities and $\sigma$ values (e.g. \citealt{Bouche2013}). 
Large line widths might also be due to turbulence in the interstellar medium induced by the interaction of nuclear activity with the host environment, as observed in low-redshift type-2 active galaxies (e.g. \citealt{Woo2016,Fischer2018,Venturi2021,Girdhar2022}). Yet, such turbulence-induced effects are often associated with relatively low centroid velocities (i.e. $|v|<200-300$ km s$^{-1}$), typical of gas at systemic or following rotation \citep{Woo2016,Fischer2018}. This does not seem to be the case of the broad components reproducing [O\,{\sc iii}] line wings in SUPER type-2 systems, which exhibit centroid velocities $|v|\gtrsim300$ km s$^{-1}$ (see green $v$ values shown in Figs. \ref{fig:super_spectra_XID}--\ref{fig:super_spectra_CID1057}), unlikely associated with galaxy rotation. Moreover, we point out that turbulence-induced effects have been traced so far in observations of local Seyfert galaxies at high spatial resolution and sensitivity (see e.g. \citealt{Fischer2018,Venturi2021,Girdhar2022}). As a consequence, it might be hard to detect them in lower-quality observations at $z\sim2$ like SUPER, with a spatial resolution of 2--4 kpc.

Similarly, the observed large [O\,{\sc iii}] line widths and prominent asymmetries (mostly blueshifted wings) can be hardly accounted for by merging scenarios. This is also supported by the lack of clear [O\,{\sc iii}] double peaks and by the relatively unperturbed morphology of stellar continuum emission, as traced by rest-frame 260 $\mu$m FIR continuum emission (when available for SUPER targets, see \citealt{Lamperti2021}). Although harder to detect than line emission due to gas, stellar continuum can better unveil the presence of a second merging galaxy. The only SUPER galaxy showing some degree of perturbed morphology is cid\_1143, which exhibits an offset of about 2 kpc between the centroids of rest-frame optical line emission (i.e. H$\alpha$ and [O\,{\sc iii}]) and rest-frame 260 $\mu$m FIR continuum, respectively \citep{Lamperti2021}. This may indicate the presence of a companion for this galaxy but, even if it did, given the offset of the possible companion from FIR continuum, this is not expected to contribute to the flux in the galaxy center, where we observe $v_{10}$ velocities of about $1000$ km s$^{-1}$. For the other sources we can instead likely rule out early mergers based on the observational evidence available so far, although the S/N of our SINFONI observations might not be sufficient to detect faint merger signatures and to identify especially final-state mergers.

On the other side, $v_{\rm 50}$ velocities are useful to describe the velocity field of the component dominating the overall kinematics of ionised gas. In all galaxies, we find moderate $v_{\rm 50}$ values ($|v_{\rm 50}|<400$ km s$^{-1}$; even $|v_{\rm 50}|<100$ km s$^{-1}$ in XID614 and cid\_2682), with a predominance of blueshifted $v_{\rm 50}$ values. This means that a significant fraction of [O\,{\sc iii}] line emission is found in the blue line wing, further supporting the presence of fast ionised outflows in these galaxies. In XID36 instead, $v_{\rm 50}$ velocities appear redshifted almost everywhere (reaching 400 km s$^{-1}$), as a consequence of the extended [O\,{\sc iii}] red wing, which dominates the overall [O\,{\sc iii}] kinematics in this object (see Fig. \ref{fig:super_spectra_XID}). In a few galaxies, we also detect a smooth velocity gradient of $v_{\rm 50}$: in cid\_1143 it is outflow-dominated, with all blueshifted $v_{\rm 50}$ velocities of increasing absolute value from NW to SE (N is up, E to left); in contrast, in XID419 and XID614 outflow emission does not seem to be the dominant component, being the $v_{\rm 50}$ field consistent with disk rotation, with velocities ranging from negative to positive values passing through the galaxy centre. The presence of outflows in these two galaxies is better highlighted by the high values of $v_{\rm 10}$ and $v_{\rm 90}$, as previously discussed.

In all maps of Figs. \ref{fig:super_maps_XID}-\ref{fig:super_maps_CID1057}, we draw two circles corresponding: to the aperture used in Sect. \ref{sec:super_integrated} to extract integrated SINFONI spectra (dashed); and to the mean \textit{H}-band PSF (dotted), namely, with a radius equal to $\langle \theta_{\rm PSF}\rangle/2$ (see Table \ref{tab:super_obs}). In XID419 (Fig. \ref{fig:super_maps_XID}) the two circles have the same radius. Compared to the corresponding \textit{H}-band PSF, we see that the extent of S/N$>$3 [O\,{\sc iii}] line emission at $|v|>300$ km s$^{-1}$ is clearly larger than the instrumental PSF radius in all galaxies except for XID614. This ensures that the high-velocity [O\,{\sc iii}] emission associated with the outflows is spatially resolved in at least six objects of our sample.
In XID614, S/N$>$3 [O\,{\sc iii}] maps and \textit{H}-band PSF have approximately the same extent in all directions
(see Fig. \ref{fig:super_maps_2}). However, the fact that we see structures and variations in gas kinematics across the FoV of XID614 suggests that high-velocity [O\,{\sc iii}] line emission at large distance is at least marginally spatially resolved in our data.

\subsection{Ionised outflow properties and energetics}\label{sec:super_energetics}

Since H$\beta$ is detected at low S/N in our \textit{H}-band observations, due to extinction effects, we derive ionised outflow properties from higher-S/N [O\,{\sc iii}] line emission\footnote{Using [O\,{\sc iii}] line emission requires more assumptions than H$\alpha$ or H$\beta$, but the latter are overall detected at a lower S/N compared to [O\,{\sc iii}], thus motivating our choice.}. Properties such as outflow velocity $v_{\rm out}$, radial extent $R_{\rm out}$, and [O\,{\sc iii}] luminosity $L^{\rm [OIII]}_{\rm out}$ can be directly measured from SINFONI data, while other physical quantities (i.e. electron density $n_{\rm e}$, oxygen abundance [O/H]) must be assumed in the calculation of outflow mass rate $\dot M_{\rm out}$.

\subsubsection{Directly measured outflow properties}\label{sec:super_dirprop}

The outflow properties we can measure directly are $v_{\rm out}$, $R_{\rm out}$ and $L^{\rm [OIII]}_{\rm out}$. As outflow velocity $v_{\rm out}$, we consider for each target the maximum absolute value of all observed $v_{10}$ or $v_{90}$ velocities (i.e. $v_{\rm out}={\rm max}[|v_{10}|,|v_{90}|]$). This definition is widely adopted in the literature (e.g. \citealt{Cresci2015,Carniani2015,Tozzi2021,Vayner2021b,Vayner2021a}) and relies on the assumption that all observed lower velocities are consequence of projection effects. Other works instead use $w_{80}$ as outflow velocity (e.g. \citealt{Harrison2012,Kakkad2016,Kakkad2020}), being $w_{80}$ less affected by projection effects compared to $v_{10}$ and $v_{90}$.
Yet, we point out that $w_{80}$ values depend more strongly on line shapes, systematically leading to higher velocities for symmetric line profiles than asymmetric ones, where only one wing is detected as a possible consequence of dust extinction. Dust effects indeed seem to be the most likely explanation for the (blueshifted) asymmetric [O\,{\sc iii}] profiles found in most of our type-2 AGN, which could be hardly accounted for by standard unified models (e.g. \citealt{Antonucci1993,Urry1995}; further discussions are presented in Sect. \ref{sec6}). 

For this reason, we here adopt $v_{\rm out}={\rm max}[|v_{10}|,|v_{90}|]$ to define the outflow velocity, unlike the previous $w_{80}$-based approach adopted in \citet{Kakkad2020} for the SUPER type-1 sample. However, as pointed out in Sect. \ref{sec:super_kinematics}, all type-2 AGN show $w_{80}>800$ km s$^{-1}$ in most of the FoV, thus meeting the $w_{80}>600$ km s$^{-1}$ criterion used in \citet{Kakkad2020} to identify [O\,{\sc iii}] outflow emission (e.g. see also \citealt{Harrison2016}).

Overall, we find wind velocities within the range $v_{\rm out}\sim600-1100$ km s$^{-1}$, with a mean outflow velocity $\langle v_{\rm out}\rangle\sim830$ km s$^{-1}$. For comparison, we also compute for each object the outflow velocity defined as $v_{\rm max}=v_{\rm bro}+2\sigma_{\rm bro}$ \citep{Rupke2013}, where $v_{\rm bro}$ and $\sigma_{\rm bro}$ are the velocity and velocity dispersion, respectively, of the broad Gaussian components associated with outflows. This parametric definition of outflow velocity typically delivers values slightly larger compared to the $v_{\rm out}$ definition based on $v_{10,90}$ but, being adopted in several works (e.g. \citealt{Brusa2015a,Fiore2017,Leung2019,Perrotta2019,Kakkad2020}), it is a useful quantity to compute for comparing our results with other from the literature (see Sect. \ref{sec6}). For XID419, we estimate $v_{\rm max}$ as twice the $\sigma$ value of the single Gaussian employed, as done in \citet{Kakkad2020}. The $v_{\rm out}$ values inferred for each galaxy are listed in Table \ref{tab:super_wind}, along with maximum observed values of $v_{\rm max}$ and $w_{80}$ (as well as other outflow properties inferred in the following).

To estimate $R_{\rm out}$ instead, we first infer the observed (maximum) outflow radius $R_{\rm obs}$, measured as the maximum extent from the centre of [O\,{\sc iii}] line emission at $|v|>300$ km s$^{-1}$. We then correct $R_{\rm obs}$ for the \textit{H}-band PSF (i.e. $R_{\rm out}= \sqrt{R^2_{\rm obs}-(\langle\theta_{\rm PSF}\rangle/2)^2}$, with $\langle\theta_{\rm PSF}\rangle$ being the average PSF FWHM, see Table \ref{tab:super_obs}), thus obtaining intrinsic radii of a few kpc for spatially resolved outflows ($R_{\rm out}\sim$2--4 kpc). For XID614, where the ionised outflow is marginally resolved, we consider the resulting value as an upper limit to $R_{\rm out}$ (i.e. $R_{\rm out}<1.9$ kpc). All $R_{\rm out}$ estimates and upper limits are listed in Table \ref{tab:super_wind}.

For a measurement of the [O\,{\sc iii}] flux associated with ionised winds, we follow the same prescriptions adopted by \citet{Kakkad2020} for the SUPER type-1 AGN. We hence consider the total [O\,{\sc iii}] flux contained in the high-velocity line channels (i.e. $|v|>300$ km s$^{-1}$), by summing up the flux contributions of all spaxels with S/N$>$3 on [O\,{\sc iii}]. Considering the quality of our data, a non-parametric approach is more suitable than a parametric one based on the results from the multi-Gaussian modelling, where the detection of a broad ‘outflow' component depends on the S/N of the data. In XID419 indeed, due to the low S/N, the [O\,{\sc iii}] line profile is sufficiently well reproduced by a single Gaussian component across the FoV, with no need for an additional component. 
Yet, the $f_{300}$ ratio map shown in Fig. \ref{fig:super_maps_XID} unveils a non-negligible fraction (varying within 0.3--0.5) of [O\,{\sc iii}] flux in $|v|>300$ km s$^{-1}$ line channels, overall suggesting that high-velocity gas is outflowing in this galaxy as well (as discussed in Sect. \ref{sec:super_kinematics}).
However, we point out that by summing spaxel-by-spaxel the flux of broad Gaussian components (for the six objects requiring a multi-Gaussian modelling of [O\,{\sc iii}]) we obtain [O\,{\sc iii}] outflow fluxes close to our non-parametric estimates, differing on average by a factor of about 2 (same found in \citealt{Kakkad2020} for type-1 AGN).

Finally, we correct all inferred (non-parametric) [O\,{\sc iii}] outflow fluxes for dust extinction using the $A_V$ values inferred in Sect. \ref{sec:super_integrated} (see Table \ref{tab:super_wind}), and convert them to intrinsic [O\,{\sc iii}] outflow luminosities $L^{\rm [OIII]}_{\rm out}$. All resulting dust-corrected $L^{\rm [OIII]}_{\rm out}$ values are also reported in Table \ref{tab:super_wind}.

\begin{table*}
\footnotesize
\centering
\caption{Main properties of ionised outflows in the seven SUPER type-2 AGN examined in this work. From left, columns list for each galaxy: (1) $V$-band extinction $A_V$ (inferred in Sect. \ref{sec:super_integrated}); (2) outflow radius $R_{\rm out}$; (3) outflow velocity $v_{\rm out}={\rm max}[|v_{10}|,|v_{90}|]$; (4) dust-corrected {\normalfont [O\,{\sc iii}]} outflow luminosity $L^{\rm [OIII]}_{\rm out}$, inferred from $|v|>300$ km s$^{-1}$ line channels; (5) outflow mass rate $\dot M_{\rm out}$; (6) escape velocity $v_{\rm esc}$ of the total galaxy system, computed at $R_{\rm out}$ (see Sect. \ref{sec:super_vesc}); maximum observed values of (7) $w_{80}$ and (8) $v_{\rm max}=v_{\rm bro}+2\sigma_{\rm bro}$ \citep{Rupke2013}.}
\begin{tabular}{lc|cccc|ccc}
\hline
Target ID & $A_V$ & $R_{\rm out}$ & $v_{\rm out}$ & $L^{\rm [OIII]}_{\rm out}$ & $\dot M_{\rm out}$ & $v_{\rm esc}$ & $w_{80}$ & $v_{\rm max}$\\
 &  & kpc & km s$^{-1}$ & 10$^{42}$ erg s$^{-1}$ & M$_{\odot}$ yr$^{-1}$ & km s$^{-1}$ & km s$^{-1}$ & km s$^{-1}$\\
 & (1) & (2) & (3) & (4) & (5) & (6) & (7) & (8)\\
\hline
XID36 & $1.3\pm0.8$ & 3.5 & $990\pm50$ & $14\pm5$ & $10\pm6$ & $860\pm220$ & $1760\pm120$ & $1450\pm80$ \\
XID419 & $>0$ & $3.6$ & $790\pm120$ & $>0.3$ & $>0.2$ & $980\pm210$ & $1020\pm110$ & $800\pm40$ \\
XID614 & $2.8\pm1.6$ & $<1.9$ & $590\pm60$ & $2.2\pm1.6$ & $>1.7$ & $>980$ & $960\pm140$ & $910\pm80$ \\
cid\_1057 & $1.8\pm1.0$ & 3.1 & $720\pm70$ & $2.2\pm1.0$ & $1.3\pm0.9$ & $960\pm240$ & $1130\pm130$ & $1420\pm130$ \\
cid\_1143 & $1.2^{+1.8}_{-1.2}$ & 2.9 & $1070\pm120$ & $3^{+14}_{-2}$ & $3^{+29}_{-2}$ & $760\pm290$ & $1320\pm70$ & $1370\pm120$ \\
cid\_2682 & $>0.6$ & 2.0 & $670\pm80$ & $>0.9$ & $>0.8$ & $1190\pm270$ & $1130\pm90$ & $1240\pm130$ \\
cid\_451 & $>2.7$ & 2.6 & $950\pm40$ & $>30$ & $>27$ & $1400\pm330$ & $1270\pm70$ & $1290\pm40$ \\
\hline
\label{tab:super_wind}
\end{tabular}
\end{table*}

\subsubsection{Ionised outflow mass rate}\label{sec:super_mdot}

 The three quantities directly measured in the previous subsection - namely, $v_{\rm out}$, $R_{\rm out}$ and $L^{\rm [OIII]}_{\rm out}$ - are fundamental ingredients to compute the mass rate of ionised outflows. Following the prescriptions adopted by \citet{Kakkad2020} for the SUPER type-1 AGN sample, we calculate ionised outflow masses $M_{\rm out}$ as follows (for an electron temperature of $T\sim10^4$ K; e.g. \citealt{Carniani2015,Kakkad2016}):

\begin{equation}
M_{\rm out}=0.8\times10^8~\bigg(  \frac{L^{\rm [O III]}_{\rm out}}{10^{44}~\text{erg s}^{-1}} \bigg)~\bigg(  \frac{n_{\rm e}}{500~\text{cm}^{-3}} \bigg)^{-1}~\frac{\text{\(\textup{M}_\odot\)}}{10^{\rm [O/H]}},
\label{eq:super_mass}
\end{equation}
\\
where all oxygen is assumed to be ionised to {\sc O}$^{2+}$, $n_{\rm e}$ and [O/H] are electron density and oxygen abundance in solar units, respectively. For a uniformly filled (i.e. constant $n_{\rm e}$) outflow of bi-conical geometry, the mass rate $\dot M_{\rm out}$ at a certain radius $R_{\rm out}$ can be then calculated as \citep{Fiore2017}: 

\begin{equation}
\dot M_{\rm out}\ =\ 3\ \frac{M_{\rm out}v_{\rm out}}{R_{\rm out}},
\label{eq:super_mdot}
\end{equation}
\\

which gives the instantaneous mass rate of ionised gas crossing a spherical sector at a distance $R_{\rm out}$ from the central AGN.

Equation \ref{eq:super_mass} requires values of [O/H] and $n_{\rm e}$, but none of them can be measured from our data. In principle, $n_{\rm e}$ can be inferred from [S\,{\sc ii}]$\lambda$6716/[S\,{\sc ii}]$\lambda$6731 flux ratio (e.g. \citealt{Osterbrock2006}), but [S\,{\sc ii}] is undetected in our SINFONI \textit{K}-band data, except for some faint [S\,{\sc ii}] emission in XID36 (see Fig. \ref{fig:super_spectra_XID}). Therefore, we assume a solar [O/H] abundance and $n_{\rm e}=500\pm250$ cm$^{-3}$ (i.e. 50\% uncertainty) in agreement with previous studies (e.g. \citealt{Storchi-Bergmann2010,Carniani2015,Riffel2015,Davies2020,Cresci2023}) and consistently with prescriptions adopted in \citet{Kakkad2020} for SUPER type-1 AGN. 

In Table \ref{tab:super_wind}, we summarise the properties of the ionised outflows in the seven SUPER type-2 AGN, with corresponding uncertainties. We also list the escape velocity $v_{\rm esc}$ from the total galaxy gravitational potential, computed at $R_{\rm out}$ (see Sect. \ref{sec:super_vesc}) and the $A_V$ estimates (or lower limits, derived in Sect. \ref{sec:super_integrated}) used to obtain the corresponding dust-corrected $L^{\rm [OIII]}_{\rm out}$ measurements (or lower limits). Errors $v_{\rm out}$, $w_{80}$ and $v_{\rm max}$ are the uncertainties resulting from the kinematic analysis, while those on $A_V$, $L^{\rm [OIII]}_{\rm out}$ and $\dot M_{\rm out}$ are computed via error propagation. Due to the asymmetric uncertainty on $A_V$ for cid\_1143, errors on the $L^{\rm [OIII]}_{\rm out}$ and $\dot M_{\rm out}$ for this object correspond to the minimum-maximum range of possible values, determined by the minimum/maximum values of the physical quantities they depend on. Despite being not obvious how to estimate an uncertainty on $R_{\rm out}$, we do not expect errors smaller than the SINFONI spatial pixel (i.e. 0.05$''$), which corresponds to about 0.4 kpc at $z\sim2$. Nevertheless, the uncertainty on $\dot M_{\rm out}$ is dominated by the large errors on $A_V$ (larger than 50\%) and $n_{\rm e}$ (assumed equal to 50\%).
Finally, we notice that the inferred upper limit to $R_{\rm out}$ in XID614 leads to a corresponding lower limit to $\dot M_{\rm out}$ (i.e. $\dot M_{\rm out}>1.7$ M$_{\odot}$ yr$^{-1}$).

Before SUPER, the galaxy cid\_415 was observed with SINFONI in AO-mode, but with the largest 0.250$''$ pixel-scale and the lower spectral-resolution \textit{HK} filter ($R\sim$1500). We notice that our inferred values of $R_{\rm out}$ and $v_{\rm out}$ ($R_{\rm out}\sim2.6$ kpc and $v_{\rm out}\sim950$ km s$^{-1}$) are both smaller than those previously found ($R_{\rm out}\sim5$ kpc and $v_{\rm out}\sim1600$ km s$^{-1}$; \citealt{Perna2015b}), based on these first SINFONI dataset. The smaller value of $R_{\rm out}$ is likely consequence of a surface brightness loss in our SINFONI data, due to the smaller pixel (0.100$''$) adopted in SUPER. Instead, the discrepancy in $v_{\rm out}$ is due to a more extreme definition of outflow velocity adopted in \citet{Perna2015b} (i.e. $v_{05}$), combined also with slightly different inferred redshifts ($z=2.444$ in this work against $z=2.45$ in \citet{Perna2015b}), which leads to a systematic difference of about $\sim$520 km s$^{-1}$ in the adopted systemic galaxy velocity (with respect to which all velocities are computed). After accounting for these discrepancies, the two $v_{\rm out}$ values are consistent.

\section{Comparison with SUPER type-1 AGN}\label{sec5}

In addition to presenting results on ionised outflows in SUPER type-2 AGN, this paper investigates ionised wind properties as a function of AGN properties (i.e. AGN bolometric luminosity $L_{\rm bol}$, hydrogen column density $N_{\rm H}$) and host galaxy parameters (i.e. stellar mass $M_*$, star formation rate, escape velocity $v_{\rm esc}$) for the full SUPER sample, aiming at unveiling any difference between type-1 and type-2 systems which possibly sheds light on the intrinsic nature of these two AGN classes. Except for $v_{\rm esc}$ (see Sect. \ref{sec:super_vesc}) and $N_{\rm H}$ (measured from archival X-ray spectra, \citealt{Circosta2018}), all other AGN/host galaxy properties were derived by \citet{Circosta2018} using the SED fitting code CIGALE (Code Investigating GALaxy Emission; \citealt{Noll2009}), and are overall confirmed by the latest updates presented in \citet{Bertola2024}. Values of $M_*$, star formation rate (SFR), $L_{\rm bol}$ and $N_{\rm H}$ for the SUPER type-2 AGN are found listed in Table \ref{tab:super_ty2sample} (see \citealt{Kakkad2020} for the type-1 AGN).

\subsection{Faster outflows in type-2 AGN at the low-luminosity end}\label{sec:super_key}

\begin{figure*}
    \centering    \includegraphics[width=0.95\linewidth]{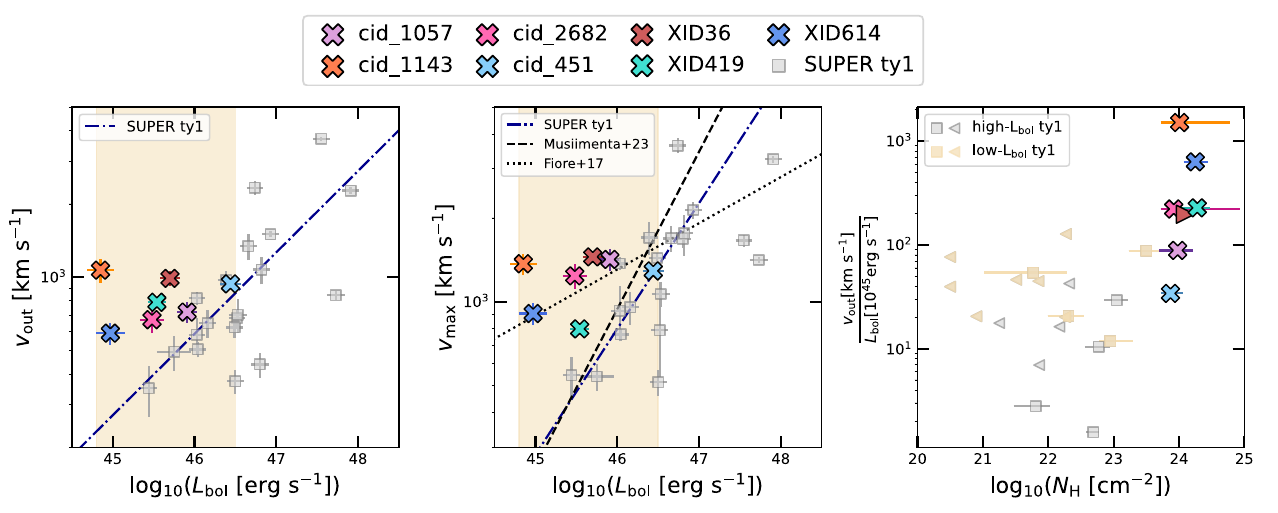}
    \caption{Outflow velocities $v_{\rm out}$ and $v_{\rm max}$ (left and middle panels) as a function of $L_{\rm bol}$, and $v_{\rm out}/L_{\rm bol}$ ratios versus X-ray $N_{\rm H}$ measurements (right panel). Coloured crosses and grey squares represent SUPER type-2 and type-1 AGN; triangles in the right panel indicate upper/lower limits to $N_{\rm H}$. Blue dashed-dotted lines represent our best-fit linear relation to the type-1 measurements, from which type-2 points depart increasingly at decreasing $L_{\rm bol}$ (mean deviation of 0.4dex). In the middle panel, we draw for comparison the empirical relation by \citet{Fiore2017} (dotted line), along with a more recent version from \citet{Musiimenta2023} (dashed). The orange shading in the left and middle panels marks the low-$L_{\rm bol}$ range (i.e. $L_{\rm bol}=10^{44.8-46.5}$ erg s$^{-1}$), where we compare type-1 and type-2 AGN measurements. In the right-hand panel we consistently use light-orange markers to identify type-1 AGN lying in the low-$L_{\rm bol}$ regime. All panels overall point to faster winds in type-2 obscured AGN than in type-1 systems at lower $L_{\rm bol}$, as a consequence of a more efficient acceleration of outflows in dusty, obscured environments.}
    \label{fig:super_voutlbol}
\end{figure*}

The key result of this work is shown by the three panels of Fig. \ref{fig:super_voutlbol}, which overall display outflow velocity as a function of AGN bolometric luminosity $L_{\rm bol}$ for the full SUPER sample, divided into type-2 (coloured thick crosses) and type-1 (grey squares) AGN. In particular, in the left and middle panels $v_{\rm out}=\textrm{max}[|v_{10}|,|v_{90}|]$ and $v_{\rm max}=v_{\rm bro}+2\sigma_{\rm bro}$ are displayed along the y-axis, respectively, with the type-1 AGN measurements taken consistently from \citet{Kakkad2020}. In addition to showing the well-known trend of larger outflow velocity at increasing AGN luminosity (e.g. \citealt{Bae2014,Woo2016,Perna2017,Rakshit2018}; discussed in Sect. \ref{sec6}), both panels unveil consistently the same following result in the lower $L_{\rm bol}$ regime (i.e. $L_{\rm bol}\sim10^{44.8-46.5}$ erg s$^{-1}$; orange shading), where all SUPER type-2 AGN are found. Interestingly, the SUPER type-2 AGN host overall faster ionised outflows than their type-1 counterparts within the same $L_{\rm bol}$ range, and appear to depart from the global $v_{\rm out,max}-L_{\rm bol}$ increasing trend followed by the SUPER type-1 AGN. Not surprisingly, the $v_{\rm max}$ value of XID419 is the lowest of the type-2 sample, since $v_{\rm max}$ for this target is computed as $2\sigma$ (see Sect. \ref{sec:super_dirprop}), with $\sigma$ referring to the single Gaussian component employed. 

To better appreciate the deviation of the type-2 velocity measurements from the type-1 ones, we fit a linear relation to the type-1 distribution in both $v_{\rm out}-L_{\rm bol}$ and $v_{\rm max}-L_{\rm bol}$ planes (blue dashed-dotted line), from which type-2 velocities deviate increasingly at lower luminosities by up to about 0.8dex in both panels, and with a mean deviation of 0.4 dex. In the middle panel, we also draw for comparison the empirical $v_{\rm max}-L_{\rm bol}$ relation inferred in \citet{Fiore2017} (black dotted line), and an updated version by \citet{Musiimenta2023} (black dashed line), including recent results and considering $z>0.5$ AGN only. The type-1 population appears to overall follow the empirical relation by \citet{Musiimenta2023}, as opposed to the type-2 measurements. We point out that the small number (i.e. seven) of type-2 measurements prevents us from an accurate linear fit to the type-2 distribution of measurements.

In Appendix \ref{sec:super_mdotlbol} we show additional plots where we compare other outflow properties and AGN/host galaxy parameters, none of which pointing to either a clear trend or interesting discrepancy between the type-1 and type-2 populations in SUPER. In particular, the left and right panels of Fig. \ref{fig:super_mstar_sfr} display $v_{\rm out}$ plotted against $M_*$ and SFR, respectively; whereas in Fig. \ref{fig:super_mdotlbol} we searched for trends of outflow mass rate with $L_{\rm bol}$. Although the $v_{\rm out}-L_{\rm bol}$ plane reveals a clear type-1/type-2 dichotomy, it is hard to establish whether a separation between the two AGN populations is also present in terms of outflows mass rate, due to the large uncertainty on this quantity. In fact, mass rate estimates strongly depend on the accuracy of other measured parameters (e.g. $A_V$, $R_{\rm out}$), and also require several quantities to be assumed (e.g. $n_{\rm e}$, [O/H], ionisation). All this uncertainty inevitably increases the scatter in the resulting mass rate values, thus possibly hiding the type-1/type-2 dichotomy instead visible in terms of the more directly inferred outflow velocities.

We investigate the potential effect of different nuclear obscuring conditions on driving outflows in the right panel of Fig. \ref{fig:super_voutlbol}, aiming at shedding light on the nature of our unveiled type-1/type-2 dichotomy. To remove any dependence of $v_{\rm out}$ on AGN luminosity (discussed in Sect. \ref{sec6}), we normalise $v_{\rm out}$ to $L_{\rm bol}$ (in units of 10$^{45}$ erg s$^{-1}$), and plot the resulting $v_{\rm out}/L_{\rm bol}$ ratios against the hydrogen column density $N_{\rm H}$, measured from X-ray data in \citealt{Circosta2018} for the full SUPER sample. Light-orange markers identify type-1 AGN lying in the low-$L_{\rm bol}$ regime (i.e. $L_{\rm bol}\sim10^{44.8-46.5}$ erg s$^{-1}$, orange shading in the left and middle panels), whereas high-$L_{\rm bol}$ type-1 objects are plotted as grey markers. Upper/lower limits to $N_{\rm H}$ are finally shown as triangles. In the right panel, we can clearly see that SUPER type-1 and type-2 AGN occupy two distinct regions of the $(v_{\rm out}/L_{\rm bol})-N_{\rm H}$ plane. The distribution of SUPER measurements along the \textit{x}-axis shows their X-ray obscured/unobscured classification (see \citealt{Circosta2018}). All SUPER type-2 AGN lie in the X-ray highly-obscured regime ($N_{\rm H}\gtrsim10^{23}$ cm$^{-2}$), whereas most of the type-1 population exhibits unobscured conditions ($N_{\rm H}\lesssim10^{22}$ cm$^{-2}$), with some system lying in the intermediate-obscuration region ($N_{\rm H}\sim10^{22}-10^{23}$ cm$^{-2}$). The combined resulting separation along the \textit{y}-axis is more interesting and unexpected: the SUPER type-2 sample indeed features $v_{\rm out}/L_{\rm bol}$ values varying within $\sim$30--1500 (with only cid\_451 smaller than 80), with a mean value of $\sim$400, while almost all type-1 AGN have ratios smaller than 50, with an overall mean value of $\sim$40. This points to the picture where AGN-driven outflows are accelerated more efficiently in more obscured environments (further discussion in Sect. \ref{sec6}).

\begin{figure}
    \centering    \includegraphics[width=0.95\linewidth]{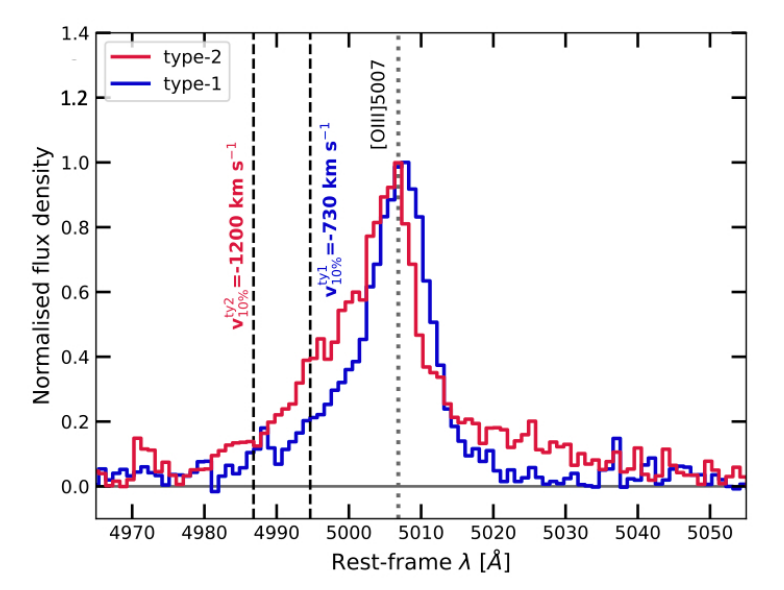}
    \caption{Stacked {\normalfont [O\,{\sc iii}]$\lambda$5007} spectra of type-2 (red) and type-1 (blue) AGN from SUPER with $L_{\rm bol}\sim10^{44.8-46.5}$ (orange shading in Fig. \ref{fig:super_voutlbol}), resulting from integrated (aperture of 0.25$''$-radius), subtracted {\normalfont [O\,{\sc iii}]} spectra with S/N$>$5. Integrated spectra are averaged by the inverse of squared noise, and then the resulting stacked spectrum is normalised to the final {\normalfont [O\,{\sc iii}]} peak. Comparing the two stacked spectra, we find a statistically significant ($2.7\sigma$) difference in {\normalfont [O\,{\sc iii}]} blue wing, which is more prominent in type-2 spectrum of about 500 km s$^{-1}$ compared to the type-1 stack.}
    \label{fig:super_stack}
\end{figure}

To further check and quantify the observed discrepancy in outflow velocity between SUPER type-1 and type-2 AGN, we directly compare the ‘mean' [O\,{\sc iii}] line profiles of SUPER type-1 and type-2 AGN, by separately stacking [O\,{\sc iii}] integrated spectra of the type-1/type-2 objects with $L_{\rm bol}\sim10^{44.8-46.5}$ erg s$^{-1}$ (Fig. \ref{fig:super_stack}): these are all seven type-2 AGN and eleven type-1 systems (i.e. those lying within the orange shading of Fig. \ref{fig:super_voutlbol}). Selecting objects over this broad (about two orders of magnitude) range of luminosity ensures us to account for possible over/underestimates of $L_{\rm bol}$ from SED fitting in type-2 AGN \citep{Circosta2018}, due to fainter AGN emission hard to constrain. However, we check the reliability of the SED-based measurements for the type-2 AGN by estimating $L_{\rm bol}$ also from dust-corrected [O\,{\sc iii}] luminosity \citep{Lamastra2009} and from X-ray 2-10 keV luminosity \citep{Duras2020}. We thus find $L_{\rm bol}$ values derived from SED fitting to agree with the [O\,{\sc iii}]-based estimates within a factor of 2 on average (computed for the four objects with $A_V$ estimates, see Table \ref{tab:super_wind}), and from X-ray-based values by a mean factor of 4, without obvious biases towards higher or lower $L_{\rm bol}$ with respect to the SED fitting derivation.

For each selected AGN, we extract an integrated [O\,{\sc iii}]$\lambda$5007 spectrum using an aperture of 0.25$''$-radius, after removing continuum emission in SUPER type-2 datacubes (see Sect. \ref{sec:super_specfit}), and continuum+BLR emission in type-1 datacubes \citep{Kakkad2020}. We exclude from the stack spectra with a S/N$<$5 on [O\,{\sc iii}], to allow for a more accurate comparison of line profiles. We thus finally remain with six S/N$>$5 type-2 spectra (all but XID419) and six S/N$>$5 type-1 spectra (X\_N\_66\_23, X\_N\_115\_23, cid\_467, S82X1905, S82X1940, S82X2058). Each type-1/type-2 spectrum is weighed by the inverse of its squared noise before being added to the stack, and then the final stacked type-2 and type-1 spectra (red and blue spectra in Fig. \ref{fig:super_stack}, respectively) are normalised to the resulting respective [O\,{\sc iii}] peak. We point out that results similar to those described below are obtained even by stacking unweighted spectra. In Fig. \ref{fig:super_stack}, we show the resulting stacked [O\,{\sc iii}] line profiles of the selected type-2 (red) and type-1 (blue) SUPER AGN, using a binning twice larger than SINFONI spectral channel to remove noisier channels. The type-2 [O\,{\sc iii}] line profile exhibits more prominent blue and red wings than type-1 spectrum. Via multiple stacks of type-2 integrated spectra, each time excluding one object, we check that all type-2 objects contribute to the [O\,{\sc iii}] blue wing, whereas the red wing is dominated by XID36. Because of the brighter and more prominent blue wing, the type-2 [O\,{\sc iii}] line profile also appears more asymmetric than the type-1 one.

We measure the $v_{10}$ velocity of each resulting stacked type-1/type-2 spectrum (vertical dashed lines in Fig. \ref{fig:super_stack}) and quantify the corresponding uncertainty as the standard deviation of the $v_{10}$ distribution resulting from a bootstrap of 400 stacks of six type-1/type-2 spectra, randomly selected with equal probability and allowing for repetitions. We thus obtain $v^{\rm ty2}_{10}=-1200\pm80$ km s$^{-1}$ for the type-2 line profile, and $v^{\rm ty1}_{10}=-730\pm90$ km s$^{-1}$ for the type-1 spectrum (see Fig. \ref{fig:super_stack}), corresponding to a $2.7\sigma$ difference in the [O\,{\sc iii}] blue wing of the AGN samples. We notice that the identification of extended [O\,{\sc iii}] wings at high velocity is generally more challenging in type-1 AGN than in type-2 systems, due to the presence of bright BLR emission (i.e. H$\beta$ and Fe\,{\sc ii}) in the  [O\,{\sc iii}] wavelength range. Therefore, as a check of the goodness of the BLR and continuum modelling of the SUPER type-1 AGN, we also stack type-1 spectra without removing BLR emission (i.e. only continuum emission removed), employing best-fit results from \citet{Kakkad2020}. We find that no evident Fe\,{\sc ii} emission bumps are present in correspondence of the [O\,{\sc iii}] line wings, and that the bulk of BLR emission in this spectral region is due to extremely large red wings of the H$\beta$ BLR line component (see Figs. A.1-A.4 in \citealt{Kakkad2020}). This supports the reliability of our comparison between type-1 and type-2 stacked spectra (see Fig. \ref{fig:super_stack}) and furthermore strengthens our result of a type-1/type-2 dichotomy in ionised outflow velocity, as firstly pointed out by our spatially-resolved measurements (shown in Fig. \ref{fig:super_voutlbol}). The origin of such an observed difference, revealing faster ionised outflows in type-2 AGN, is discussed in Sect. \ref{sec6}.

\subsection{Can ionised outflows escape galaxy gravitational potential?}\label{sec:super_vesc}

\begin{figure*}
    \centering    \includegraphics[width=0.7\linewidth]{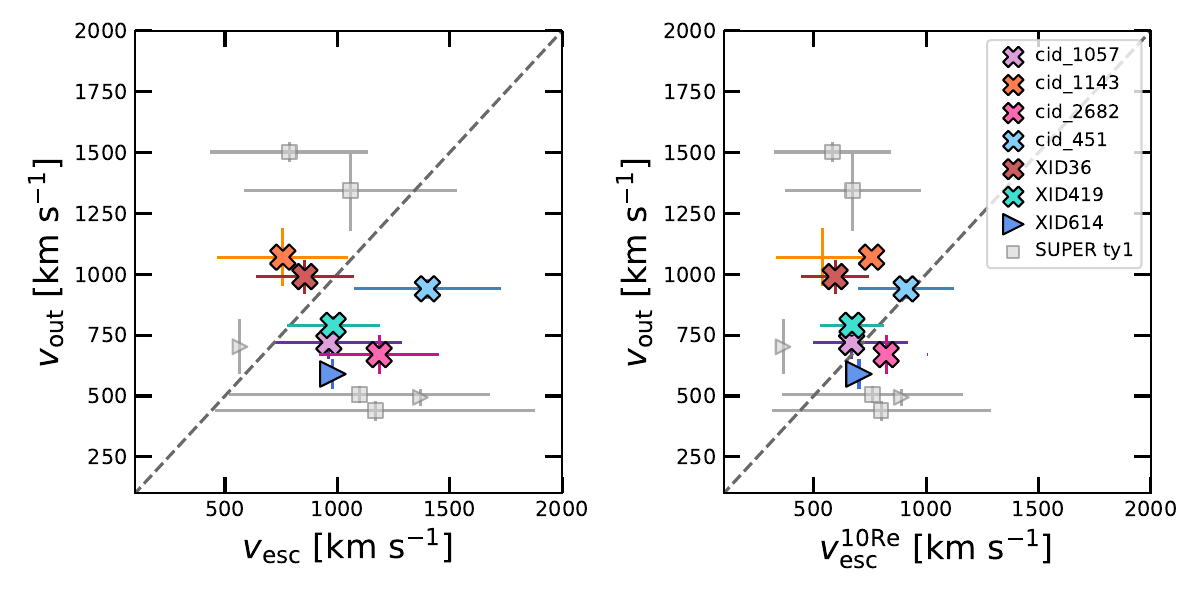}
    \caption{Outflow velocity $v_{\rm out}$ as a function of escape velocity computed at $R_{\rm out}$, as inferred from our mass modelling of SUPER type-2 (coloured symbols) and all those type-1 (grey) AGN galaxies with known $M_*$, hence not $L_{\rm bol}$-matched to the type-2 AGN sample. We display two sets of escape speed values: one inferred according to the classical definition of escape velocity to infinity ($v_{\rm out}$, left panel), hence that one needed to escape the DM halo ($R_{\rm vir}\sim80-160$ kpc); the other one corresponding to the escape velocity required to reach $10~R_{\rm e}$ ($v^{\rm 10Re}_{\rm esc}$, right panel), that is $\sim$30--50 kpc for the galaxies here considered. Triangles represent lower limits to the escape speed (due to an inferred upper limit to $R_{\rm out}$), whereas dashed line indicates $v_{\rm out}=v_{\rm esc}$ and $v_{\rm out}=v^{\rm 10Re}_{\rm esc}$ values, respectively. The overall distribution of measurements suggests that, although possibly unable to escape the DM halo, the outflows however can impact the host galaxy reservoir up to galaxy scales (i.e. tens of kpc).}
    \label{fig:super_vesc}
\end{figure*}

One of the main channels via which outflows can quench (or at least reduce) galactic SF is through the ejection of a substantial amount of gas out of the host galaxy, otherwise available to form new stars. This requires outflows energetic enough to efficiently sweep away the gas on large scales, eventually escaping the gravitational potential of their galaxy.
\citet{Kakkad2020} presented a first investigation of the ‘ejective' ability of the ionised outflows detected in the SUPER type-1 sample and found that, for most galaxies, only a modest fraction ($<10$\%) of the outflowing gas can effectively escape the host's gravitational potential well. In this section, we compare the outflow velocities with the escape velocity of each galaxy for the full SUPER sample, homogeneously for all objects using a more accurate mass modelling. 

Following prescriptions from the literature \citep{Marasco2023}, we use the \texttt{python} package \texttt{galpy} \citep{Bovy2015} to build a dynamical mass model consisting of a dark matter (DM) halo, a stellar disk, and a gaseous disk, to derive the total escape velocity profile for SUPER targets with known $M_*$, derived from SED fitting \citep{Circosta2018}. All examined type-2 AGN galaxies have $M_*$ estimates ranging within $M_*\sim10^{10.7-11.2}$ M$_{\odot}$ (see Table \ref{tab:super_ty2sample}), while for only few type-1 systems $M_*$ estimates are available ($M_*\sim10^{10.1-11.2}$ M$_{\odot}$). In fact, stellar emission is typically buried in the bright AGN component of type-1 systems, thus preventing us from constraining $M_*$ in most cases (only six SUPER type-1; see Table A.2 in  \citealt{Circosta2018}).

For the DM halo, we assume a Navarro-Frenk-White (NFW, \citealt{Navarro1997}) profile, with a virial mass $M_{200}$ derived from $M_*$ via the stellar-to-halo mass relation of \citet{Girelli2020}, and a concentration $c$ determined from the $M_{200}-c$ relation of \citet{Dutton2014}; both relations are computed at the redshift of each target. We model the stellar disk with a double-exponential profile with scale-length $R_{\rm d}$ and scale-height $R_{\rm d}$/5, assuming $R_{\rm d}$ equal to the half-light radius $R_{50}$ divided by 1.68 (correct for a pure exponential disk, Sérsic index $n=1$). We determine $R_{50}$ from the size-$M_*$ relation for a disk galaxy ($n=1$) at z$\sim2.2$ \citep{Mowla2019}. Since we have CO-based measurements of molecular gas mass $M_{\rm gas}$ for a few SUPER AGN \citep{Circosta2021}, we model a gaseous disk of $M_{\rm gas}$ equal to the mean value measured for SUPER targets (i.e. $M_{\rm gas}\sim6\times10^9$ M$_{\odot}$), and same size as the stellar disk, since we expect the molecular gas to extend on scales comparable to those of stars. In doing so, we derive the escape velocity profile of the total modelled mass distribution (i.e. DM halo + stellar disk + gas disk).

Figure \ref{fig:super_vesc} displays $v_{\rm out}$ as a function of the escape velocity at the maximum inferred outflow radius $R_{\rm out}$ for SUPER type-2 (coloured symbols) and all those type-1 (grey) AGN with known $M_*$ \citep{Circosta2018}, hence not $L_{\rm bol}$-matched to the type-2 AGN sample. The two panels display data points corresponding to two distinct definitions of escape speed: one derived according to the classical definition\footnote{The minimum velocity required by an object to reach to infinity, with a final null velocity.} ($v_{\rm esc}$ in the left panel; also listed in Table \ref{tab:super_wind}), which translates in more practical terms to the velocity needed to reach the virial radius $R_{\rm vir}$ of the DM halo, as derived from our mass modelling ($R_{\rm vir}\sim80-160$ kpc); the other one corresponding to a less rigid definition of escape speed ($v^{\rm 10Re}_{\rm esc}$ in the right panel), namely the velocity required to reach $10~R_{\rm e}$ (i.e. $\sim$30-50 kpc, depending on the galaxy). This second set of values shows whether the outflows can travel up to distances of tens of kpc from the centre of the galaxy, and escape (at least) the gravitational potential of the galaxy baryonic component (i.e. stellar disk + gas disk). Triangles indicate lower limits to the escape speed, corresponding to upper limits to $R_{\rm out}$, since escape speed curves increase at decreasing $R_{\rm out}$. A dashed line indicates the locus of points where $v_{\rm out}$ are equal to escape speed values. Under the simplifying assumption that our dynamical modelling introduces no further uncertainty, we consider the error on our escape speed values to be due to the total uncertainty on $M_*$, including both statistical (from SED fitting \citealt{Circosta2018}) and systematic ($\sim$0.1dex; \citealt{Pacifici2023}) uncertainties.

 In the left panel of Fig. \ref{fig:super_vesc}, the overall majority of galaxies have $v_{\rm out}$ values consistent or nearly consistent with $v_{\rm esc}$ to infinity. Unfortunately, the uncertainties and simplified assumptions of our dynamical modelling prevent us from firmly testing whether these outflows can effectively sweep up a significant amount of gas out of the DM halo, as predicted by the ejective feedback scenario. Indeed, we derived the outflow velocity (assumed to be constant in our model) from our observed velocity measurements, which are subject to inclination effects and limited to few kpc from the centre of each galaxy, beyond which line emission gets fainter and the S/N drops. Moreover, our dynamical model includes only gravity and no hydrodynamic effects, which may instead play a major role. Outflows could hence get trapped in gaseous halos, eventually cool and be re-accreted, even if they have high initial velocities. This would contribute to hampering the wind ejective ability to expel the gas reservoir of their host, unlike model predictions (e.g. \citealt{King2015,Zinger2020}). Yet, the distribution of measurements in the right panel of Fig. \ref{fig:super_vesc} strongly suggests that, even if these outflows do not successfully escape the total gravitational potential well (i.e. the DM halo), they are likely to impact the host gas reservoir up to galaxy scales (i.e. 30--50 kpc). They can hence sweep gas out of the galaxy disk and/or provide feedback through a ‘preventive' mode (e.g. \citealt{VanDeVoort2011}), by injecting energy and driving turbulence in the surrounding medium. All this will eventually prevent gas from cooling and collapsing to form new stars. Even inferring escape speeds at a fixed radius of 50 kpc for all galaxies, we obtain escape speeds similar to $v^{\rm 10Re}_{\rm esc}$. Finally, as opposed to Fig. \ref{fig:super_voutlbol}, we notice that Fig. \ref{fig:super_vesc} shows no clear separation between type-1 and type-2 AGN measurements.
\section{Discussion}\label{sec6}

Our analysis has unveiled that SUPER type-2 AGN host faster ionised outflows than their type-1 counterparts in the lower $L_{\rm bol}$ regime ($L_{\rm bol}\sim10^{44.8-46.5}$), as traced by higher [O\,{\sc iii}] velocities. At low redshift ($z<0.8$), several studies have searched for differences in ionised outflow properties between type-2 and type-1 systems in large AGN samples (e.g. \citealt{Mullaney2013,Bae2014,Rakshit2018,Wang2018,Rojas2020}). In general, they found narrower and more symmetric [O\,{\sc iii}] line profiles in type-2 AGN, indicating the detection of both blueshifted and redshifted outflow components; whereas, almost exclusively blueshifted outflows in type-1 AGN, with typically larger line widths and more negative velocities. This can be explained in terms of biconical outflow geometry in the context of standard unification models (e.g. \citealt{Antonucci1993,Urry1995}), as a consequence of orientation effects along the line of sight, combined with dust obscuration. Indeed, assuming a nearly face-on orientation of type-1 AGN, with well visible BLR emission along the line of sight, the receding cone of the outflow remains hidden on the the opposite side of the galaxy disk plane.

However, most of these low-redshift studies do not consider the well known dependence of outflow kinematics on AGN luminosity and accretion rate (e.g. \citealt{Bae2014,Woo2016,Rakshit2018}), and overall compare more luminous type-1 AGN with fainter type-2 samples, possibly leading to an only apparent discrepancy. Moreover, nowadays observational evidence increasingly challenges the standard unified picture of AGN, such as the discovery of changing-look AGN (see \citealt{Ricci2023} for a recent review), which favours the interpretation of the type-1/type-2 AGN dichotomy in terms of distinct phases within a common evolutionary sequence.

\begin{figure*}
    \centering    \includegraphics[width=0.95\linewidth]{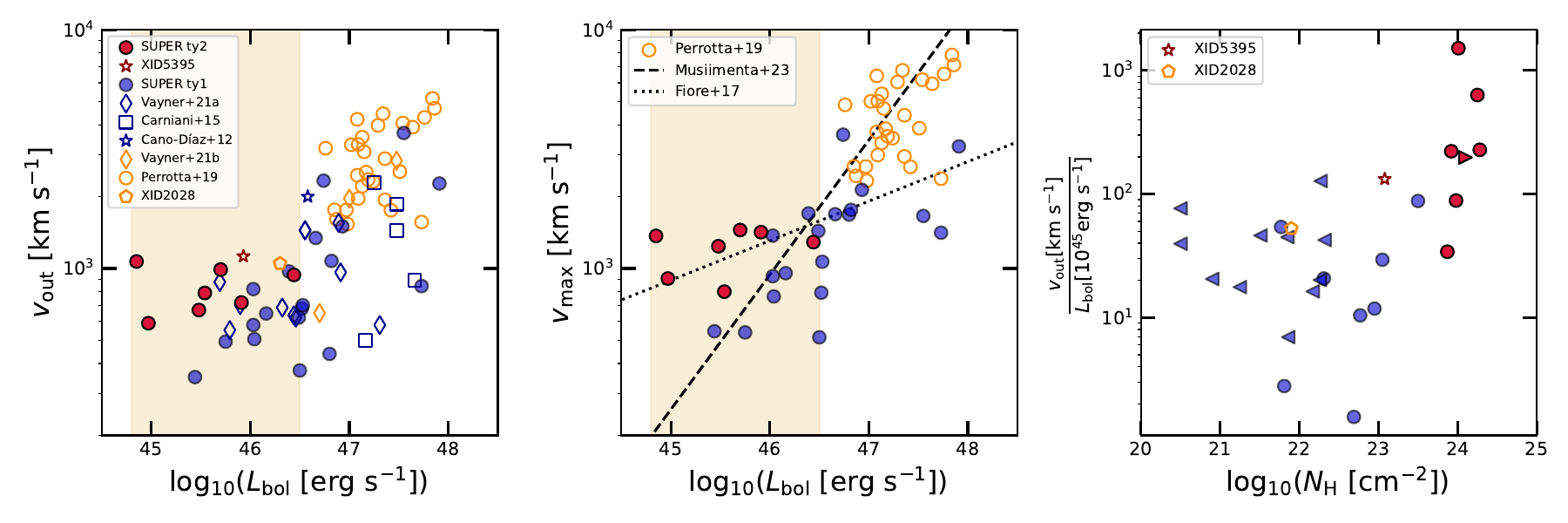}
    \caption{Same panels as in Fig. \ref{fig:super_voutlbol}, showing SUPER type-2 (red filled circles) and type-1 (blue filled circles) measurements along with IFS-based results from the literature of other type-2 (red empty star; XID5395, \citealt{Brusa2016}) and type-1 AGN (blue empty symbols; \citealt{CanoDiaz2012,Carniani2015,Vayner2021b}) at $z\sim1-3$. Orange symbols represent red obscured systems which are nevertheless optically classified as type-1 AGN \citep{Perrotta2019,Vayner2021a,Cresci2023,Veilleux2023}. Outflow velocities from \citet{Brusa2016} and \citet{Perrotta2019} have been corrected to account for discrepancies with $v_{\rm out}$ and $v_{\rm max}$ definitions here adopted (see Sect. \ref{sec6}). While ‘standard' blue type-1 and red type-2 from the literature agree with values found in SUPER, the orange points, almost all at $L_{\rm bol}\gtrsim10^{47}$ erg s$^{-1}$, clearly show that higher outflow velocities are recovered if we select optical type-1 but dusty obscured objects.}
    \label{fig:super_voutlbol_literature}
\end{figure*}

At high redshift ($z>1$), where more complex processes are expected to take place - especially at crucial cosmic epochs of galaxy evolution such as $z\sim2$ - orientation effects might play a minor role compared to low redshift. Previous works have indeed provided evidence that at high redshift the bulk of the dust obscuration comes from the host galaxy rather than being due to the classical dusty torus (e.g. \citealt{Gilli2014,Gilli2022,Perna2018,Circosta2019}). This is consistent with results from numerical simulations according to which different AGN classes correspond to distinct evolutionary phases of galaxies’ life cycle \citep{Hopkins2006,Menci2008}. In particular, X-ray bright but obscured ($L_{\rm 2-10~keV}>10^{44}$ erg s$^{-1}$, $N_{\rm H}\gtrsim10^{22}$ cm$^{-2}$) AGN at $z\sim2$ have been proposed to be in the active ‘blow-out’ phase of AGN galaxies’ evolution (e.g. \citealt{Brusa2010,Urrutia2012}) and, in fact, powerful outflows have been detected in such systems (e.g. \citealt{Brusa2015a,Brusa2016,Zakamska2016,Perna2015a,Perna2015b,Vayner2023,Glikman2024}).
In these obscured sources, dusty material favours the acceleration of outflows (e.g. \citealt{Ishibashi2015,Costa2018}), which efficiently clear out obscuring dust and finally unveil the AGN. Radiation-pressure on dust has been indeed identified as the main mechanism responsible for the larger [O\,{\sc iii}] velocities and line widths observed in extremely red and luminous quasars (ERQs; $L_{\rm bol}>10^{47}$ erg s$^{-1}$), compared to blue quasars at $z=2-3$ \citep{Perrotta2019}.

However, \citet{Villar2020} provided a revised comparison of the [O\,{\sc iii}] kinematics in 20 ERQs from \citet{Perrotta2019} with other AGN samples matched in luminosity, and found the [O\,{\sc iii}] velocities of ERQs to be overall consistent with those of luminous blue quasars. A similar result was also found by \citet{Temple2019}, who compared the [O\,{\sc iii}] kinematics in a sample of luminous ($L_{\rm bol}\approx10^{47}$ erg s$^{-1}$), heavily reddened quasars at $z>2$ to unobscured quasars in the same luminosity and redshft range. On the contrary, larger [O\,{\sc iii}] velocities have been observed by \citet{DiPompeo2018} in obscured AGN at $z<0.4$, resembling the obscured high-redshift population based on their optical/IR colours, and matched in luminosity with unobscured objects. All these controversial results show how complex carrying out such a comparison is, and leaves the debate open.

Figure \ref{fig:super_voutlbol_literature} displays the same panels as in Fig. \ref{fig:super_voutlbol} and compares the results obtained for SUPER targets with measurements from the literature. In particular, we collect from the literature only results based on IFS observations and with outflow velocities defined consistently with ours in SUPER, to allow for a proper comparison as homogeneous as possible. In all three panels of Fig. \ref{fig:super_voutlbol_literature}, SUPER results of type-2 and type-1 AGN are shown as red and blue filled markers, respectively, along with other IFS-based measurements of type-2 (red empty star; XID5395, \citealt{Brusa2016}) and type-1 (blue empty symbols; \citealt{CanoDiaz2012,Carniani2015,Vayner2021b}) AGN at $z=1-3$ from the literature. In orange, we highlight the measurements relative to red obscured AGN showing BLR emission (hence, optical type-1 systems), which should be considered as an intermediate class of objects between standard (blue) type-1 and (red) type-2 AGN. These are: the obscured quasar XID2028 at $z\sim1.6$ (orange empty pentagon; \citealt{Cresci2023,Veilleux2023}), the powerful obscured ERQs from \citet{Perrotta2019} (orange empty circles); and the sample of obscured quasars from \citet{Vayner2021a} (orange diamonds), among which all but one (the lowest-velocity point) are selected as ERQs. 


The first panel compares $v_{\rm out}$ measurements from SUPER and the literature based on $v_{10}$/$v_{90}$, as obtained by \citet{Vayner2021b,Vayner2021a} (blue/orange diamonds) and by \citet{Cresci2023} and \citet{Veilleux2023} consistently for XID2028; whereas higher percentile outflow velocities were used in \citet{Brusa2016} and \citet{Perrotta2019} (i.e. $v_{05}$ and $v_{02}$, respectively). Therefore, we correct outflow velocities inferred in \citet{Brusa2016} and \citet{Perrotta2019} by a factor of 0.86 and 0.77, respectively, corresponding to the mean ratio of $v_{10,90}$-to-$v_{05,95}$ and $v_{10,90}$-to-$v_{02,98}$ measured for SUPER type-2 AGN. We also add measurements from \citet{CanoDiaz2012} and \citet{Carniani2015}. Similarly, we re-scale $v_{02}$ outflow velocities from \citet{Perrotta2019} by a factor of 1.2, as obtained from the mean $v_{{\rm max}}$-to-$v_{02,98}$ ratio for SUPER type-2 AGN, and plot them in the $v_{{\rm max}}-L_{\rm bol}$ plane (middle panel), along with $v_{{\rm max}}$ measurements for SUPER type-2 (derived in Sect. \ref{sec:super_dirprop}) and type-1 AGN \citep{Kakkad2020}.

As shown in Fig. \ref{fig:super_voutlbol_literature}, standard blue (unobscured) type-1 and red (obscured) type-2 AGN from the literature lie close to SUPER type-1 and type-2 points, respectively. For instance, the obscured type-2 AGN XID5395 (empty red star; \citealt{Brusa2016}) exhibits outflow velocity and $v_{\rm out}/L_{\rm bol}$ ratio (i.e. $\sim$130) comparable with those of SUPER type-2 systems, and a column density ($N_{H}\sim10^{23}$ cm$^{-2}$) that places this source in the nuclear highly-obscured regime. Within the optical type-1 AGN population instead, we notice that ‘hybrid' sources (i.e optical type-1 but X-ray obscured AGN; orange markers) show higher outflow velocities, consistently with the picture proposed in this work. In fact, the hybrid type-1 AGN XID2028 (empty pentagon; \citealt{Perna2015a,Cresci2023,Veilleux2023}) behaves like the SUPER type-2 objects in terms of $v_{\rm out}-L_{\rm bol}$, while it is more similar to SUPER type-1 counterparts in the $v_{\rm out}/L_{\rm bol}-N_{\rm H}$ parameter space, falling at the boundary of the obscured vs. unobscured regimes ($N_{H}\sim10^{22}$ cm$^{-2}$).

Therefore, radiation pressure of dust might be the dominant mechanism explaining the presence of faster-ionised outflows in type-2 AGN, also at lower bolometric luminosity with respect to that typical of ERQs ($L_{\rm bol} > 10^{47}$ erg s$^{-1}$) from \citet{Perrotta2019}. Our analysis has furthermore revealed a clear type-1/type-2 dichotomy which fully reflects the X-ray unobscured/obscured classification based on obscuring material on nuclear scales. Interestingly, SUPER type-2 AGN are all X-ray highly obscured with $N_{\rm H}\gtrsim10^{23}$ cm$^{-2}$, which may explain the higher observed velocities of their ionised outflows, compared to those observed in SUPER type-1 AGN within the same luminosity range. Rather than being a consequence of projection effects, this result supports the scenario in which obscured optical type-2 AGN at $z\sim2$ might indeed evolve towards the unobscured type-1 phase, as radiation-pressure driven outflows blow out obscuring material and unveil the central AGN and its BLR emission.

Despite previous studies already pointed to individual sources as likely ‘blow-out' candidates (e.g. \citealt{Brusa2016}), SUPER has provided observational evidence of such an evolutionary scenario in a larger AGN sample, and also pointed to an optical type-1/type-2 AGN dichotomy in outflow velocity, for the first time, in a sample of moderate-luminous AGN at $z\sim2$ (i.e. $L_{\rm bol}\sim10^{44.8-46.5}$ erg s$^{-1}$). As a comparison, the ERQ sample at $z=2-3$ studied by \citet{Perrotta2019} consists of highly luminous AGN ($L_{\rm bol}>10^{47}$ erg s$^{-1}$), all showing BLR emission (hence, optical type-1 systems). On another side, the published results from the KASHz survey \citep{Harrison2016} relative to X-ray-selected AGN ($L_{\rm 2-10~keV}>10^{42}$ erg s$^{-1}$) at $z=0.6-1.7$, shows no evidence for more extreme gas kinematics in obscured (type-2) objects than in unobscured (type-1) counterparts (see left panel of Fig. \ref{fig:super_voutlbol}). We point out that this is likely consequence of the lower S/N of KASHz data compared to SUPER, which hampers the detection of any extended [O\,{\sc iii}] line wing. This is indeed the case for cid\_1143 and cid\_451, among the targets in common to both surveys (Scholtz et al., in prep.), where the shallower KMOS data completely miss the broad [O\,{\sc iii}] wing due to the outflow, thus underestimating its velocity.

\section{Conclusions}\label{sec7}
In this paper we presented SINFONI \textit{H}- and \textit{K}-band observations of the SUPER type-2 AGN sample, thus complementing the results previously on the type-1 systems \citep{Kakkad2020,Kakkad2023}. The spectral analysis of SINFONI datacubes (Sect. \ref{sec:super_specfit}) allowed us to study the spatial distribution and kinematics of ionised gas via [O\,{\sc iii}] line emission (Sect. \ref{sec:super_kinematics}), and found evidence for ionised outflows in all seven examined type-2 AGN, moving at $v_{\rm out}>600$ km s$^{-1}$. The detected outflows are spatially resolved in six objects and extend up to 2--4 kpc from the centre, while marginally resolved in the remaining source (XID614, $R_{\rm out}<1.9$ kpc). We also compute outflow mass rates from dust-corrected [O\,{\sc iii}] outflow luminosity (Sect. \ref{sec:super_mdot}).

In Sect. \ref{sec5}, we investigated outflows velocity as a function of AGN bolometric luminosity, aiming at unveiling any difference between the type-1 and type-2 AGN populations in SUPER. The key result of this paper is presented in Sect. \ref{sec:super_key} and clearly shown in Fig. \ref{fig:super_voutlbol}, where the outflow velocity is plotted against $L_{\rm bol}$: SUPER type-2 AGN host faster winds than the type-1 systems within the same luminosity range (i.e. $L_{\rm bol} = 10^{44.8-46.5}$ erg s$^{-1}$). Interestingly, these type-2 AGN (namely, all type-2 systems with high-S/N [O\,{\sc iii}] detection in our SINFONI observations) are all highly obscured systems ($N_{\rm H}\gtrsim10^{23}$ cm$^{-2}$). By separately stacking integrated type-1 and type-2 spectra, we found a 2.7$\sigma$ difference in the blue wing of the [O\,{\sc iii}] line profile, which appears to be more prominent in type-2 AGN than in type-1 counterparts (Fig. \ref{fig:super_stack}). We interpret this as a consequence of the larger amount of obscuring material ($N_{\rm H}\gtrsim10^{24}$ cm$^{-2}$ in all examined type-2 AGN), which favours the acceleration of winds via radiation-pressure on dust (e.g. \citealt{Ishibashi2015,Costa2018}). 

SUPER survey has hence provided observational evidence of an evolutionary origin of the type-1/type-2 dichotomy at $z\sim2$, consistently with theoretical predictions and with previous observational evidence, whereas orientation effects might play a minor role. Within this picture, type-2 AGN - even at lower luminosity - represent the blow-out phase (e.g. \citealt{Hopkins2008}), whose obscured environment favours the acceleration of fast winds via radiation pressure on dust. Such an interpretation was already proposed for obscured AGN at $z\sim2$ (e.g. \citealt{Brusa2015a,Brusa2016,Perna2015b,Zakamska2016}), but not clearly associated with the type-2 optical class, with some of these examined X-ray obscured AGN being optical type-1 systems (e.g. \citealt{Perna2015a,Perrotta2019}). Moreover, the ionised outflows overall detected in SUPER AGN galaxies have velocities compatible with the escape speed of their DM halo (Fig. \ref{fig:super_vesc}, left panel), pointing to no type-1/type-2 discrepancy though. Unfortunately, the uncertainties and the simplified assumptions of our dynamical model prevent us from testing definitely the ejective feedback scenario. However, we found that, although possibly unable to successfully escape the DM halo of their galaxy, these ionised outflows are likely to reach 30--50 kpc scales (Fig. \ref{fig:super_vesc}, right panel), where they can still provide effective feedback by sweeping away gas and/or injecting energy in the galaxy medium (‘preventive' mode), thus halting the formation of new stars.

To confirm this novel result found in SUPER, in future we will need larger samples of high-redshift type-1 and type-2 AGN. The new generation of IFS facilities, both from space (e.g NIRSpec on \textit{JWST}) and on the ground with AO (e.g. ERIS at VLT), will allow us to exploit as much as possible the power of spatially resolved observations, to finally take the next step to fully understand feedback mechanisms and AGN evolution at the Cosmic Noon.

\begin{acknowledgements}
      
      We thank Giacomo Venturi for the insightful discussion and his advise on dynamical mass modelling. We acknowledge financial support from INAF under the Large Grant 2022 “The metal circle: a new sharp view of the baryon cycle up to Cosmic Dawn with the latest generation IFU facilities”. We also acknowledge support from the Italian Ministry for University and Research (MIUR) for the BLACKOUT project funded through grant PRIN 2017PH3WAT.003, and for the project “Black Hole winds and the Baryon Life Cycle of Galaxies: the stone-guest at the galaxy evolution supper”, funded through grant PRIN \#2017PH3WAT. MP acknowledges support from Grant PID2021-127718NB-I00 funded by the Spanish Ministry of Science and Innovation/State Agency of Research (MICIN/AEI/ 10.13039/501100011033). CMH acknowledges funding from a United Kingdom Research and Innovation grant (code: MR/V022830/1). IL acknowledges support from PID2022-140483NB-C22 funded by AEI 10.13039/501100011033 and BDC 20221289 funded by MCIU by the Recovery, Transformation and Resilience Plan from the Spanish State, and by NextGenerationEU from the European Union through the Recovery and Resilience Facility. AP acknowledges support by an Anniversary Fellowship at the University of Southampton.
\end{acknowledgements}

%
\bibliographystyle{aa} 
\bibliography{biblio} 
%

\begin{appendix}

\section{Further comparison of outflow properties with AGN/host parameters}\label{sec:super_mdotlbol}

\begin{figure*}
    \centering    \includegraphics[width=0.6\linewidth]{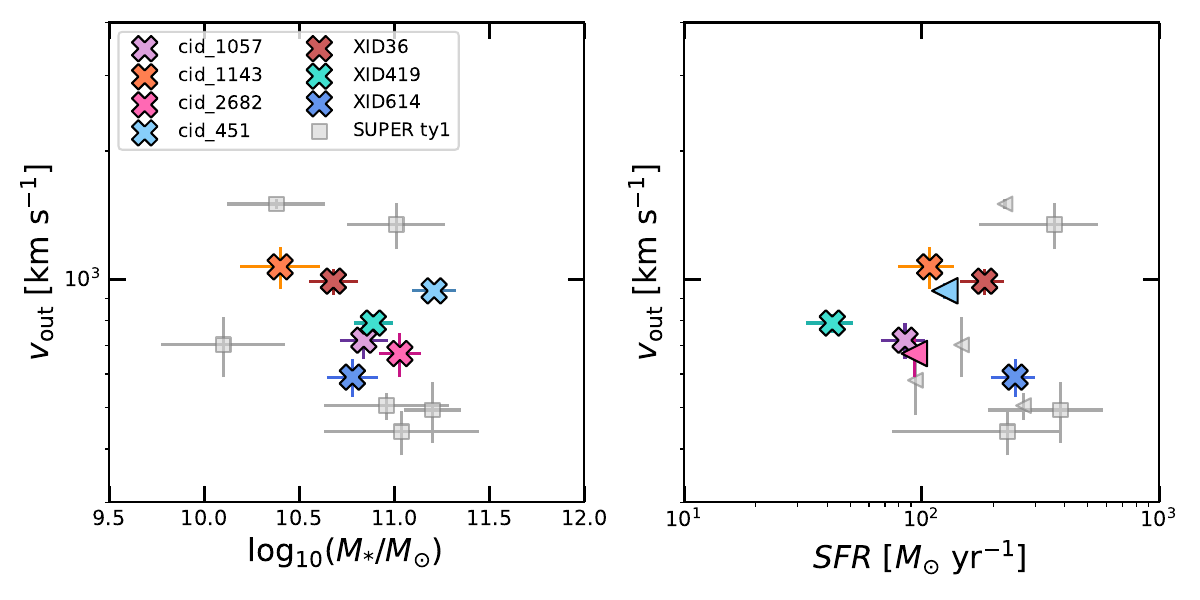}
    \caption{Outflow velocity $v_{\rm out}$ as a function of host galaxy stellar stellar mass ($M_*$, left panel) and star formation rate (SFR, right panel), for those SUPER AGN sources with available SED-based estimates or upper limits \citep{Circosta2018}. Errorbars on $M_*$ and SFR values represent the total uncertainty, including both statistical and typical systematic uncertainties ($\sim$0.1dex in $M_*$ and 0.3dex in SFR, \citealt{Pacifici2023}). Colours and symbols are the same as in Figs. \ref{fig:super_voutlbol} and \ref{fig:super_vesc}. The two panels show neither clear trend of $v_{\rm out}$ with the host stellar properties nor separation between type-1 and type-2 measurements, which appear scattered together across the two planes.}
    \label{fig:super_mstar_sfr}
\end{figure*}

In this appendix we further search for any dependence of outflow properties on AGN/host galaxy parameters, yet finding no result as interesting as that shown in Fig. \ref{fig:super_voutlbol}. In Fig. \ref{fig:super_mstar_sfr} we plot $v_{\rm out}$ against $M_*$ and SFR in the left and right panel, respectively, for those SUPER AGN sources with available $M_*$ and SFR estimates (or upper limits) from SED fitting \citep{Circosta2018}. As pointed in Sect. \ref{sec:super_vesc}, constraining (or at least placing upper limits to) stellar properties might be particularly challenging in bright type-1 AGN, where the dominant AGN emission outshines the fainter stellar counterpart. For this reason, whereas the seven SUPER type-2 AGN galaxies have all $M_*$ and SFR estimates (or upper limits), only six and seven type-1 systems have estimates (or upper limits) of $M_*$ and SFR available, respectively. Errors on $M_*$ and SFR estimates are the total uncertainty, including both statistical (from SED fitting) and typical systematic uncertainties ($\sim$0.1dex in $M_*$ and 0.3dex in SFR, \citealt{Pacifici2023}). Colours and symbols are the same as in Figs. \ref{fig:super_voutlbol} and \ref{fig:super_vesc}. We point out that these $M_*$ and SFR values derived by \citet{Circosta2018} are consistent within the uncertainty with the updated SED fitting results \citep{Bertola2024} obtained with CIGALE \citep{Noll2009,Yang2020}, recently implemented to fit also X-ray data. Unlike Fig. \ref{fig:super_voutlbol}, we find no clear trend of $v_{\rm out}$ with either $M_*$ or SFR, with type-1 and type-2 measurements scattered together across the two planes. All this overall suggests no connection of $v_{\rm out}$ with host galaxy stellar properties, which is instead well known to increase with AGN luminosity (e.g. \citealt{Bae2014,Woo2016,Rakshit2018}). However, we argue that the modest number of measurements and the difficulty in constraining stellar properties of AGN galaxies (especially for type-1 AGN hosts) might conceal some link between these quantities.

Considering the interesting type-1/type-2 dichotomy found by comparing outflow velocity with bolometric luminosity (see Fig. \ref{fig:super_voutlbol}), we check whether the same discrepancy between the SUPER type-2 and type-1 samples is also present in terms of outflow mass rates. To compare type-1 and type-2 measurements as homogeneously as possible, we re-compute outflow mass rates for the type-1 sample consistently with the prescriptions adopted on $v_{\rm out}$ and $R_{\rm out}$ in this work (see Sect. \ref{sec:super_energetics}), still using values inferred in \citet{Kakkad2020}. As discussed in Sect. \ref{sec:super_dirprop}, $v_{10}$ and $v_{90}$ indeed provide a more suitable definition of $v_{\rm out}$ than $w_{80}$, being less sensitive to the line shape which may be driven by dust extinction. Regarding the value of $R_{\rm out}$ to use in Eq. \ref{eq:super_mdot}, the type-1 outflow mass rates were all computed at a fixed radius of 2 kpc \citep{Kakkad2020}, since [O\,{\sc iii}] line emission is unresolved in most of type-1 observations, as opposed to type-2 datasets. Therefore, for the few type-1 AGN with clear detection of spatially resolved outflows, we consistently take as outflow radius $R_{\rm out}$ the maximum distance of [O\,{\sc iii}] outflow emission \citep{Kakkad2020}. For all other unresolved type-1 AGN, we use the maximum distance (or corresponding upper limit) from the central AGN where is found the bulk of the ionised gas, as resulted from spectroastrometry analysis \citep{Kakkad2020}. With these values of $v_{\rm out}$ and $R_{\rm out}$, we obtain $\dot M_{\rm out}$ type-1 mass rates, consistent with the type-2 estimates. 

\begin{figure*}
    \centering    \includegraphics[width=0.6\linewidth]{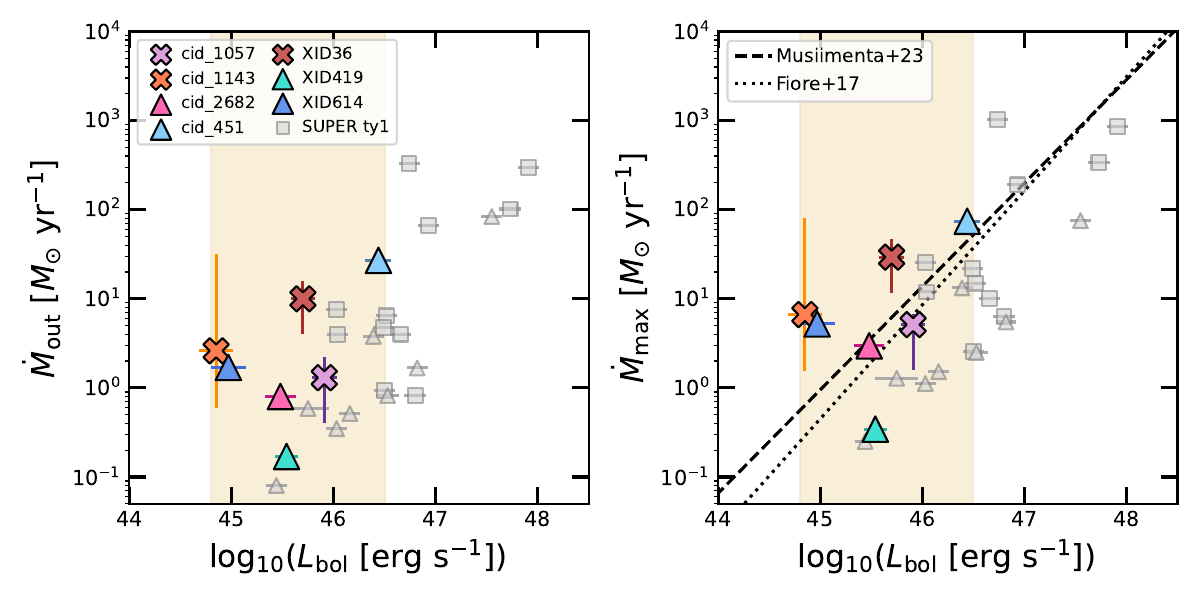}
    \caption{Outflow mass rates as a function of $L_{\rm bol}$ for the full SUPER sample. Upper and lower panels show our trust $\dot M_{\rm out}$ values and $\dot M_{\rm max}$ mass rates, derived using $v_{\rm max}$ in Eq. \ref{eq:super_mdot}. Triangles indicate lower limits to mass rates. In the lower panel, scaling relations from \citet{Fiore2017} and \citet{Musiimenta2023} are shown as dotted and dashed lines, after being re-scaled to $n_{\rm e}=500$ adopted for our targets, while values of $\dot M_{\rm max}$ have been multiplied by a factor of 2 to match definitions adopted in \citet{Fiore2017} and \citet{Musiimenta2023}. Compared to Fig. \ref{fig:super_voutlbol}, the distribution of measurements is more scattered, with no clear discrepancy between the type-1 and type-2 samples. The comparison is furthermore hampered by the large number of lower limits to mass rates.}
    \label{fig:super_mdotlbol}
\end{figure*}

In Fig. \ref{fig:super_mdotlbol}, we plot outflow mass rates as a function of $L_{\rm bol}$ for the full SUPER sample. The upper panel shows the $\dot M_{\rm out}$ values (same coloured markers as in Fig. \ref{fig:super_voutlbol}), derived according to our trust prescriptions (see Sect. \ref{sec:super_energetics}), most of which are lower limits (triangles) due to an upper limit to $R_{\rm out}$ and/or a lower limit to $A_V$ (see Table \ref{tab:super_wind}). Similarly, the lower panel shows the $\dot M_{\rm max}$ values obtained by using $v_{\rm max}$ values \citep{Rupke2005} in Eq. \ref{eq:super_mdot}, which can be compared with the scaling relations by \citet{Fiore2017} (dotted line) and \citet{Musiimenta2023} (dashed). To allow a proper comparison, both scaling relations have been re-scaled to our adopted $n_{\rm e}\sim500$ cm $^{-3}$, whereas all $\dot M_{\rm max}$ points have been corrected for a factor 2, since the scaling relations have been obtained by considering outflow mass rates computed from the flux of broad Gaussian components. The factor 2 corresponds to the approximated mean ratio of [O\,{\sc iii}] luminosity from broad Gaussian components to that from line channels at $|v|>300$ km s$^{-1}$, obtained separately for both type-2 and type-1 AGN \citep{Kakkad2020}.

As opposed to Fig. \ref{fig:super_mdotlbol}, the large number of inferred lower limits to mass rates hampers any accurate comparison within the low-luminosity range previously selected ($L_{\rm bol}\sim10^{44.8-46.5}$; orange shading). However, excluding lower limits, we find mean values of $\dot M_{\rm out}\sim 5$ M$_{\odot}$ yr$^{-1}$ and $\dot M_{\rm max}\sim7$ M$_{\odot}$ yr$^{-1}$ for type-2 AGN, and $\dot M_{\rm out}\sim5$ M$_{\odot}$ yr$^{-1}$ and $\dot M_{\rm max}\sim15$ M$_{\odot}$ yr$^{-1}$ for low-$L_{\rm bol}$ selected type-1 AGN. Compared to Fig. \ref{fig:super_voutlbol}, here it is harder to establish whether a separation between type-2 and type-1 AGN is actually present or not, due to the large uncertainty on outflow mass rates. Indeed, as explained in Sect. \ref{sec:super_mdot}, mass rate estimates require several quantities to be assumed (e.g. $n_{\rm e}$, [O/H], ionisation), and strongly depend on the accuracy of the other measured parameters (e.g. $A_V$, $R_{\rm out}$). All this uncertainty contributes to increasing the scatter in the resulting mass rate values, which might hide any possible type-1/type-2 dichotomy, visible instead in terms of outflow kinematics. Finally, we argue that the larger scatter in $\dot M_{\rm out}$ measurements further complicates any search for trends with $M_*$ and SFR estimates, which are more uncertain than $L_{\rm bol}$, as discussed at the beginning of this appendix.

\end{appendix}

\end{document}